\documentclass[11pt]{article}

\usepackage{graphics,graphicx,color,float}
\usepackage{epsfig,bbm}
\usepackage{bm}
\usepackage{mathtools}
\usepackage{amsmath}
\usepackage{amssymb}
\usepackage{amsfonts}
\usepackage{amsthm}
\usepackage {times}
\usepackage{setspace}
\usepackage[usenames,dvipsnames]{xcolor}
\usepackage{geometry}

\newcommand{\beq}{\begin{equation}}
\newcommand{\enq}{\end{equation}}
\newcommand{\bel}{\begin{lemma}}
\newcommand{\enl}{\end{lemma}}
\newcommand{\bet}{\begin{theorem}}
\newcommand{\ent}{\end{theorem}}

\newcommand{\tr}{\mathrm{Tr}}

\newcommand{\E}{\mathbb{E}}
\newcommand{\Tr}{\mathrm{Tr}}
\newcommand{\ketbra}[1]{|#1\rangle\langle#1|}

\newcommand{\eps}{\varepsilon}
\newcommand{\alice}{\ensuremath{\mathsf{Alice}}}
\newcommand{\bob}{\ensuremath{\mathsf{Bob}}}
\newcommand{\charlie}{\ensuremath{\mathsf{Charlie}}}
\newcommand{\id}{\ensuremath{\mathrm{I}}}

\newcommand*{\cC}{\mathcal{C}}

\newcommand*{\cH}{\mathcal{H}}

\newcommand*{\cB}{\mathcal{B}}
\newcommand*{\cD}{\mathcal{D}}

\newcommand*{\cN}{\mathcal{N}}
\newcommand*{\cS}{\mathcal{S}}

\newcommand*{\cE}{\mathcal{E}}

\newcommand{\suppress}[1]{}

\newcommand{\defeq}{\ensuremath{ \stackrel{\mathrm{def}}{=} }}
\newcommand{\F}{\mathrm{F}}
\newcommand{\Pur}{\mathrm{P}}
\newcommand {\br} [1] {\ensuremath{ \left( #1 \right) }}

\newcommand {\minusspace} {\: \! \!}
\newcommand {\smallspace} {\: \!}
\newcommand {\fn} [2] {\ensuremath{ #1 \minusspace \br{ #2 } }}
\newcommand {\ball} [2] {\fn{\mathcal{B}^{#1}}{#2}}
\newcommand {\relent} [2] {\fn{\mathrm{D}}{#1 \middle\| #2}}
\newcommand {\varrelent} [2] {\fn{V}{#1 \middle\| #2}}
\newcommand {\dmax} [2] {\fn{\mathrm{D}_{\max}}{#1 \middle\| #2}}

\newcommand {\dmaxeps} [3] {\fn{\mathrm{D}^{#3}_{\max}}{#1 \middle\| #2}}
\newcommand {\dmineps} [3] {\fn{\mathrm{D}^{#3}_{\mathrm{H}}}{#1 \middle\| #2}}
\newcommand {\dseps} [3] {\fn{\mathrm{D}^{#3}_{s}}{#1 \middle\| #2}}
\newcommand {\dsepsalt} [3] {\fn{\tilde{\mathrm{D}}^{#3}_{s}}{#1 \middle\| #2}}
\newcommand {\mutinf} [2] {\fn{\mathrm{I}}{#1 \smallspace : \smallspace #2}}
\newcommand {\imax}{\ensuremath{\mathrm{I}_{\max}}}
\newcommand {\imaxeps} [1] {\ensuremath{\mathrm{\tilde{I}}^{#1}_{\max}}}
\newcommand {\imaxepss} [1] {\ensuremath{\mathrm{I}^{#1}_{\max}}}
\newcommand {\imaxepsbeta} [2] {\ensuremath{\mathrm{\bar{I}}^{#1,#2}_{\max}}}
\newcommand {\condmutinf} [3] {\mutinf{#1}{#2 \smallspace \middle\vert \smallspace #3}}

\newcommand {\barq} {\bar{q}}

\newcommand{\bra}[1]{\langle #1|}
\newcommand{\ket}[1]{|#1 \rangle}

\mathchardef\mhyphen="2D

\makeatletter
\newcommand*{\rom}[1]{\expandafter\@slowromancap\romannumeral #1@}
\makeatother

\mathchardef\mhyphen="2D

\newtheorem{definition}{Definition}
\newtheorem{claim}{Claim}
\newtheorem{fact}{Fact}

\newtheorem{theorem}{Theorem}
\newtheorem{lemma}{Lemma}
\newtheorem{corollary}{Corollary}
\begin {document}
\title{One-shot entanglement assisted classical and quantum communication over noisy quantum channels: A hypothesis testing and convex split approach}

\author{
Anurag Anshu\footnote{Centre for Quantum Technologies, National University of Singapore, Singapore. \texttt{a0109169@u.nus.edu}} \qquad
Rahul Jain\footnote{Centre for Quantum Technologies, National University of Singapore and MajuLab, UMI 3654, 
Singapore. \texttt{rahul@comp.nus.edu.sg}} \qquad 
Naqueeb Ahmad Warsi\footnote{Centre for Quantum Technologies and SPMS, NTU, Singapore and IIITD, Delhi. \texttt{warsi.naqueeb@gmail.com}} 
}

\maketitle

\begin{abstract}
Capacity of a quantum channel characterizes the limits of reliable communication through a noisy quantum channel. This fundamental information theoretic question is very well studied specially in the setting of many independent uses of the channel. An important scenario, both from practical and conceptual point of view, is when the channel can be used only once. This is known as the one-shot channel coding problem. We provide a tight characterization of the one-shot entanglement assisted classical capacity of a quantum channel. We arrive at our result by introducing a  simple decoding technique which we refer to as \textit{position-based decoding}. We also consider two other important quantum network scenarios: quantum channel with a jammer and quantum broadcast channel.  For these problems, we use the recently introduced \textit{convex split technique} \cite{ADJ14} in addition to position based decoding. Our approach exhibits that the simultaneous use of these two techniques provides a uniform and conceptually simple framework for designing communication protocols for quantum networks.  
\end{abstract}
\section{Introduction}

A classical description of our world entails several limitations on what can be achieved physically. Law of conservation of energy prevents energy to be created out of nothing, thermodynamics disallows machines with efficiency beyond the Carnot's limit, inertia restrains motion when there is no force as a motive. These limitations have been so pivotal in the scientific revolution that they can now be found even in the laws of information (Landauer's principle \cite{Landauer61}, Shannon's capacity theorem \cite{Shannon}) and computation (Turing's halting theorem \cite{Turing}, P vs NP conjecture \cite{Cook71}). Their knowledge allows us to optimize our efforts as we seek the best possible results.

The theory of quantum information and computation, aided with the power of entanglement, opens up new possibilities. Bell's landmark theorem \cite{Bell64} tells us that quantum systems possess correlations that go beyond those achievable by classical means. Shor's algorithm \cite{Shor97} shows how a quantum computer can perform integer factoring exponentially faster than known classical algorithms. Quantum cryptography offers protocols which achieve information theoretic security in the task of key distribution \cite{LoChau99}. As these results begin to point to a physical reality that surpasses some well known boundaries in classical physics and computing, a fundamental technological limitation is brought upon us, quite ironically, by quantum entanglement itself. This is the limitation imposed by quantum noise.

Quantum noise, also known as a \textit{quantum channel}, describes the process by which a quantum particle (possessed by an experimenter) gets correlated or entangled with the environment (upon which experimenter has no control). This can be particularly unsuitable when two experimenters wish to send messages to each other and the intermediate channel has noisy behavior. Efforts to understand and mitigate quantum noise have largely developed on two fronts: communication through a quantum channel (starting from the work of Holevo \cite{Holevo98} and Schumacher and Westmoreland \cite{SchuW97}) and quantum error correction (starting from the work of Shor \cite{Shor95}). 

Here, we consider the case of communication through a quantum channel, and more specifically, the \textit{entanglement assisted classical capacity} of a quantum channel. Entanglement assistance, a widely used terminology for communicating with the help of entanglement shared between Alice (sender) and Bob (receiver), leads to two very important protocols in quantum information theory. Quantum teleportation \cite{Teleportation93} allows Alice to send a qubit to Bob using two bits of classical communication and superdense coding \cite{BennettW92} allows Alice to send two bits of message with one qubit. These two protocols strongly suggest that the presence of entanglement (upon which Alice and Bob have full control) can play important role in the process of reliable communication through a quantum channel.

Bennett et al. \cite{BennettSST02} characterized the  limits of classical communication over a noisy quantum channel when the sender and receiver share entanglement between them. They studied the case where Alice was allowed to use the channel arbitrarily many times and after each use, the channel had no memory of this use. In practice however there could be several issues, for example the channel between Alice and Bob may not be memoryless and Alice may even be forced to use the channel only once (this has been a driving force behind the emerging field of one-shot information theory). Often there are  more than one sender and receiver. For example, a quantum satellite may be beaming back information simultaneously to different base stations on earth, and these base stations may have no way of reliable collaboration between themselves. Sometimes the receiver may not have a complete knowledge of the channel characteristics, such as in the case of a quantum communication channel with an adversary or a jammer. 

We consider each of the scenarios mentioned above and provide a unified approach for designing communication protocols for them. We use two ingredients in our protocols: the technique of \textit{position based decoding} that we introduce for the protocol described in Figure \ref{fig:pointtopointprotocol}, and the technique of \textit{convex split} (introduced in \cite{ADJ14}, discussed in our context for the protocols described in Figures \ref{fig:sideinfoachieve} and \ref{fig:quantummartonprotocol}). Position based decoding (where the term decoding refers to the strategy performed by the receiver) allows the receiver to accomplish the task of \emph{quantum hypothesis testing}. In a communication protocol between Alice and Bob, as Alice sends messages to Bob through the channel, different quantum states are formed on Bob's side as a function of the message Alice has sent. Bob, who does not know the message, should be able to distinguish between these quantum states in order to learn the message. A simplification of this problem is the task of quantum hypothesis testing, where Bob should be able to distinguish between two possible quantum states with small error. Position based decoding allows Bob to distinguish between many possible quantum states that may arise from Alice's messages, if he is able to distinguish between two given quantum states. 

\vspace{0.1in}

\noindent{\bf Point to point case:} The first protocol we design concerns \textit{point to point} quantum channels, where there are two parties Alice (sender) and Bob (receiver). Alice, who is allowed to use the channel only once, wants to communicate message $m$ chosen with some a priori distribution from the set $[1:2^R]$ to Bob across the quantum channel $\cN_{A \to B}$ such that Bob is able to guess the correct message with probability at least $1-\eps$ ($\eps$ is a small constant). This we refer to as an $(R,\eps)$ \textit{entanglement assisted code} for the quantum channel $\cN_{A\to B}$. The goal here is to determine largest possible value of  $R$ (the amount of reliable communication in bits between Alice and Bob). Figure \ref {fig:pointtopointprotocol} gives a schematic of our protocol for this scenario. We show that the largest possible value of $R$ is quantified in terms of the \textit{hypothesis testing divergence}. Given two quantum states $\rho$ and $\sigma$, the hypothesis testing divergence $\dmineps{\rho}{\sigma}{\eps}$ captures the probability that an experimenter, who only wishes to accept $\rho$, ends up accepting $\sigma$. Formally, it is defined as $\dmineps{\rho}{\sigma}{\eps}:=\max_{\Lambda : \Tr(\Lambda\rho)\geq 1-\eps} \log \frac{1}{\Tr(\Lambda\sigma)}$, where $0 \prec\Lambda \preceq \mathbb{I}$ is a positive operator. Using this quantity, our main theorem is as follows, which is discussed in detail in Section \ref{sec:ptop}.

\vspace{0.1in}

\begin{theorem} 
\label{theo:ptopintro}
Let $\cN_{A \to B }$ be the quantum channel and let $\eps,\delta \in (0,1)$. Let $A'\equiv A$ be a purifying register. Then, for any $R$ smaller than
\begin{eqnarray*}
\max _{\ketbra{\psi}_{A A'}}&&\dmineps{\cN_{A \to B } (\ketbra{\psi}_{AA'})}{\cN_{A\to B}(\psi_{A}) \otimes \psi_{A'}}{\eps} \\ &&- 2\log \frac{1}{\delta},
\end{eqnarray*}
there exists an $(R,2\eps + 2\delta)$ entanglement assisted code for the quantum channel $\cN_{A \to B }.$
\end{theorem}


\noindent{\bf Outline of the protocol:} Fix a quantum state $\ket{\psi}_{AA'}.$ Alice and Bob share  $ 2^{\dmineps{\cN_{A \to B } (\psi_{A'A})}{\cN_{A  \to B}(\psi_{A}) \otimes \psi_{A'}  }{\eps}}$ independent copies of the state $\ket{\psi}_{A^\prime A},$ where the register $A$ is held by Alice and the register $A'$ is held by Bob. Each of these copies are uniquely assigned to a message $m \in [1:2^R].$ This assignment is known to both Alice and Bob. To send the message $m,$ Alice transmits her part of the $m$-th copy of the shared state over the channel.  Notice that at the end of this transmission the joint state between the $m$-th register of Bob and the channel output is $\cN_{A \to B } (\psi_{AA'})$ and the joint state for every other register $j \neq m$ and the channel output is $\cN_{A  \to B}(\psi_{A}) \otimes \psi_{A'}.$ Thus, if Bob is equipped with a binary measurement $\Lambda$ (obtained from the definition of $\dmineps{\cN_{A \to B } (\psi_{AA'})}{\cN_{A  \to B}(\psi_{A}) \otimes \psi_{A'}  }{\eps}$) which can differentiate the state $\cN_{A \to B } (\psi_{AA'})$ from $\cN_{A  \to B}(\psi_{A})  \otimes \psi_{A'}$, then he can design his (multiple outcome) decoding measurement as follows. His measurement operator corresponding to the outcome $m$ is $\Lambda \otimes \mathbb{I}$, where $\Lambda$ acts only on the channel output $B$ and the $m$-th copy of $A'$  and $\mathbb{I}$ is the identity operator on the rest of Bob's registers.  We term this decoding strategy as \textit{position based decoding}. Our protocol discussed above guarantees that Alice can communicate with Bob $ \max _{\ketbra{\psi}_{ AA'}}\dmineps{\cN_{A \to B } (\psi_{AA'})}{\cN_{A \to B}(\psi_{A}) \otimes \psi_{A'}  }{\eps}+O(\log(\eps))$ bits. This rate is also near optimal, owing to the converse bound shown in \cite{MatthewsW14}.

\vspace{0.1in}

\noindent{\bf Resource utilization:} The number of qubits of entanglement required in the above one-shot protocol is quite large, and in the asymptotic and i.i.d. setting it can grow exponentially in the number of channel uses. In order to reduce the number of qubits of the shared entanglement, we make two observations. First is that our one-shot protocol consumes only one copy of the shared entanglement and returns the rest with very small error. Thus, a large part of the shared entanglement serves as a \textit{catalyst}. The second observation, motivated by the work \cite{StrelchukHO13} and made precise in Theorem \ref{theo:ptopiidachieve} (Section \ref{sec:ptopasymptoticiid}) below, is that the entanglement can be efficiently consumed by encoding the messages in the sets of positions (instead of just one position). Both observations are used in Section \ref{sec:ptopasymptoticiid} to construct an appropriate asymptotic and i.i.d. version of the position-based decoding. This leads to a protocol that has the same rate of communication and the rate of required entanglement as the protocol constructed in \cite{BennettSST02}.

\vspace{0.1in} 

\noindent{\bf Gel'fand-Pinsker Channel:} Our second protocol concerns communication in the presence of a malicious jammer, where Alice is aware of this jammer, whereas Bob has no information about this jammer. This model was analyzed in the classical case by Gel'fand-Pinsker in their seminal work \cite{GelfandP80}. The formal setting in the quantum case is as follows (see, for example, \cite{Frederic10}): Alice shares an entangled state $\ket{\phi}_{S'S}$ with the channel itself, where the register $S'$ is held by Alice and the register $S$ is held by the channel. Unlike in the point to point case, the channel (represented by $\cN_{AS \to B}$) takes as input both $S$ and $A$. Alice wants to communicate message $m $ chosen from the set $[1:2^R]$ to Bob across the quantum channel $\cN_{AS\to B}.$ 
It is quite natural to expect that because of the absence of the knowledge of register $S$ at Bob's side, the value of $R$ (the amount of reliable communication in bits between Alice and Bob) will be smaller than the one achieved for the point to point channel. A schematic of our protocol for this task is presented in Figure \ref{fig:sideinfoachieve} and details appear in Section \ref{sec:gelfandpinsker}.

\vspace{0.1in} 


\noindent{\bf Outline of the Protocol:} Fix a state ${\psi}_{A'SA}$ such that ${\psi}_S = {\phi}_S.$ At the start of the protocol Alice and Bob share $2^{\dmineps{\cN_{AS \to B } (\psi_{A'SA})}{\cN_{AS \to B }(\psi_{AS}) \otimes \psi_{A'}}{\eps}}$ independent copies of the state $\ket{\psi}_{A''A'}$ where the register $A''$ is held by Alice and $A'$ by Bob ($\ket{\psi}_{A''A'}$ is a purification of $\psi_{A'}$). These copies are subdivided into bands of equal size $  2^{\imaxeps{\eps}(A:S)_{\psi_{A'SA}}}$. There is a unique band $\cB(m)$ for each message $m \in [1:2^R].$ To send the message $m$, Alice creates the state (close to) ${\psi}_{A'SA}$ in the register $A$ in her possession, the register $S$ with the jammer and a random register $A'$ in the band $\cB(m)$, using the convex split technique (along with Uhlmann's theorem) .  Alice then transmits the register $A$ over the channel.  Now, using position based decoding, Bob is able to decode the correct message with high probability. Thus,  Alice is able to communicate $\dmineps{\cN_{AS \to B } (\psi_{A'SA})}{\cN_{AS \to B }(\psi_{AS}) \otimes \psi_{A'}}{\eps} - \imaxeps{\eps}(A:S)_{\psi_{A'SA}}$ bits to Bob. 

\vspace{0.1in} 

\noindent{\bf Quantum Broadcast Channel}: The final case that we consider is that of quantum broadcast channel studied in the classical case (among others) by Marton in her seminal work \cite{Marton79}.  Here, Alice wishes to communicate message $m_1$ to Bob and message $m_2$ to Charlie simultaneously. While Bob and Charlie are  not allowed to collaborate with each other, the noisy channel may give correlated output to them, which makes the setting different from two independent cases of point to point channel. The channel $\cN_{F\rightarrow BC}$ takes input $F$ from Alice and produces outputs $B$ (with Bob) and $C$ (with Charlie). 

Our protocol for this task is again based on similar framework of using convex split technique and position based decoding. Convex split technique is used by Alice to establish an appropriate correlated state between Bob, Charlie and the channel output, following which Bob and Charlie perform position based decoding on their respective shares of this correlated state.  A schematic of our protocol is discussed in Figure~\ref{fig:quantummartonprotocol} and details appear in Section \ref{sec:broadcast}

\vspace{0.1in}

\noindent{\bf Outline of the Protocol:} Fix a state ${\psi}_{FA_1A_2}.$ At the start of the protocol Alice and Bob share $$ 2^{\dmineps{\Tr_C\cN_{F \to BC } (\psi_{FA_1})}{\Tr_C\cN_{F \to BC } (\psi_{F})\otimes \psi_{A_1}}{\eps} }$$ independent copies of the state $\ket{\psi}_{A'_1A_1}$ where the register $A'_1$ is held by Alice and $A_1$ by Bob ($\ket{\psi}_{A'_1A_1}$ is a purification of ${\psi}_{A_1}$). These copies are subdivided into bands of equal sizes, where each band is uniquely assigned to a message $m_1 \in [1:2^{R_1}].$ Similarly, Alice and Charlie share $$2^{\dmineps{\Tr_B\cN_{F \to BC } (\psi_{FA_2})}{\Tr_B\cN_{F \to BC } (\psi_{F})\otimes \psi_{A_2}}{\eps} }$$ independent copies of the state $\ket{\psi}_{A'_2A_2}$ where the register $A'_2$ is held by Alice and $A_2$ by Charlie ($\ket{\psi}_{A'_2A_2}$ is a purification of ${\psi}_{A_2}$). These copies are subdivided into bands of equal sizes where each band is uniquely assigned to a message $m_2 \in [1:2^{R_2}].$ The constraint on the band size is that for every $(m_1, m_2) \in [1:2^{R_1}] \times [1:2^{R_2}],$ we have $|\cB(m_1)| \times |\cC(m_2)| = 2^{\imaxepsbeta{\eps}{\delta}(A_1:A_2)_{\psi_{A_1A_2}}},$ where $\cB(m_1)$ is the band corresponding to the message $m_1$ and likewise $\cC(m_2)$ for the message $m_2$. To send the message pair $(m_1,m_2)$ Alice uses the convex split technique (along with Uhlmann's theorem) to prepare the state $\psi_{FA_1A_2}$, where register $F$ is held by Alice, register $A_1$ is a random register in $\cB(m_1)$ held by Bob and register $A_2$ is a random register in $\cC(m_2)$ held by Charlie. Alice transmits her share $F$ of the state $(\psi_{FA_1A_2})$ over the quantum channel $\cN_{F \to BC}.$ On receiving their respective shares of the channel output, Bob and Charlie employ the position based decoding to output their respective messages.

\subsection*{Comparision to previous works}

These tasks have been studied previously in classical and quantum one-shot and asymptotic settings. The works \cite{DattaTW2016,Frederic10, DattaH13} obtained a bound for point-to-point entanglement assisted quantum channel. However, their bounds do not match the converse result obtained in \cite{MatthewsW14}. The quantum Gel'fand-Pinsker channel and quantum broadcast channel were studied in \cite{Frederic10} where they obtained one-shot bounds different from ours (their bounds and our bounds converge in the asymptotic i.i.d case). An important feature of our one-shot bounds is that their forms bear close resemblance to the known results in the classical and {\em classical-quantum} settings, for example, for the point-to-point channel~\cite{WangR12}, broadcast channel \cite{RadhakrishnanSW16} and Gelf'and-Pinsker channel \cite{WarsiJ16, Tan2013}. Such is not the case with the bounds obtained in the aforementioned works on one-shot entanglement assisted quantum capacities. Another important point is that most of the previous works including~\cite{DattaTW2016,Frederic10} used the technique of {\em decoupling} through random unitaries to obtain their bounds, which is different from our techniques.

Classical analogues of our proof techniques of convex-split and position-based decoding have recently been presented in \cite{AnshuJW17classical}. Using these, we can obtain analogous results for classical versions of all the tasks considered in this paper. In the classical case, it is in fact possible to remove shared randomness by standard derandomization arguments (in the setting of average error for a prior distribution over the messages, instead of worst case error).

\section{Preliminaries}
\label{sec:prelim}

Consider a finite dimensional Hilbert space $\cH$ endowed with an inner product $\langle \cdot, \cdot \rangle$ (in this paper, we only consider finite dimensional Hilbert-spaces). The $\ell_1$ norm of an operator $X$ on $\cH$ is $\| X\|_1:=\Tr\sqrt{X^{\dagger}X}$ and $\ell_2$ norm is $\| X\|_2:=\sqrt{\Tr XX^{\dagger}}$. A quantum state (or a density matrix or a state) is a positive semi-definite matrix on $\cH$ with trace equal to $1$. It is called {\em pure} if and only if its rank is $1$. A sub-normalized state is a positive semi-definite matrix on $\cH$ with trace less than or equal to $1$. Let $\ket{\psi}$ be a unit vector on $\cH$, that is $\langle \psi,\psi \rangle=1$.  With some abuse of notation, we use $\psi$ to represent the state and also the density matrix $\ketbra{\psi}$, associated with $\ket{\psi}$. Given a quantum state $\rho$ on $\cH$, {\em support of $\rho$}, called $\text{supp}(\rho)$ is the subspace of $\cH$ spanned by all eigen-vectors of $\rho$ with non-zero eigenvalues.
 
A {\em quantum register} $A$ is associated with some Hilbert space $\cH_A$. Define $|A| := \dim(\cH_A)$. Let $\mathcal{L}(A)$ represent the set of all linear operators on $\cH_A$. Let $\mathcal{P}(A)$ represent the set of all positive semidefinite operators on $\cH_A$. We denote by $\mathcal{D}(A)$, the set of quantum states on the Hilbert space $\cH_A$. State $\rho$ with subscript $A$ indicates $\rho_A \in \mathcal{D}(A)$. If two registers $A,B$ are associated with the same Hilbert space, we shall represent the relation by $A\equiv B$.  Composition of two registers $A$ and $B$, denoted $AB$, is associated with Hilbert space $\cH_A \otimes \cH_B$.  For two quantum states $\rho\in \mathcal{D}(A)$ and $\sigma\in \mathcal{D}(B)$, $\rho\otimes\sigma \in \mathcal{D}(AB)$ represents the tensor product (Kronecker product) of $\rho$ and $\sigma$. The identity operator on $\cH_A$ (and associated register $A$) is denoted $\id_A$. For any operator $O$ on $\cH_A$, we denote by $\{O\}_+$ the subspace spanned by non-negative eigenvalues of $O$ and by $\{O\}_-$ the subspace spanned by negative eigenvalues of $O$.  For a positive semidefinite operator $M\in \mathcal{P}(A)$, the largest and smallest non-zero eigenvalues of $M$ are denoted by $\lambda_{max}(M)$ and $\lambda_{min}(M)$, respectively.

Let $\rho_{AB} \in \mathcal{D}(AB)$. We define
\[ \rho_{B} := \Tr_{A}\rho_{AB}
:= \sum_i (\bra{i} \otimes \id_{B})
\rho_{AB} (\ket{i} \otimes \id_{B}) , \]
where $\{\ket{i}\}_i$ is an orthonormal basis for the Hilbert space $\cH_A$.
The state $\rho_B\in \mathcal{D}(B)$ is referred to as the marginal state of $\rho_{AB}$. Unless otherwise stated, a missing register from subscript in a state will represent partial trace over that register. Given a $\rho_A\in\mathcal{D}(A)$, a {\em purification} of $\rho_A$ is a pure state $\rho_{AB}\in \mathcal{D}(AB)$ such that $\Tr{B}{\rho_{AB}}=\rho_A$. Purification of a quantum state is not unique.

A quantum {map} $\cE: \mathcal{L}(A)\rightarrow \mathcal{L}(B)$ is a completely positive and trace preserving (CPTP) linear map (mapping states in $\mathcal{D}(A)$ to states in $\mathcal{D}(B)$). A {\em unitary} operator $U_A:\cH_A \rightarrow \cH_A$ is such that $U_A^{\dagger}U_A = U_A U_A^{\dagger} = \id_A$. An {\em isometry}  $V:\cH_A \rightarrow \cH_B$ is such that $V^{\dagger}V = \id_A$ and $VV^{\dagger} = \id_B$. The set of all unitary operations on register $A$ is  denoted by $\mathcal{U}(A)$.

\begin{definition}
We shall consider the following information theoretic quantities. Reader is referred to ~\cite{Renner05, Tomamichel09,Tomamichel12,Datta09, TomHay13} for many of these definitions. We consider only normalized states in the definitions below. Let $\varepsilon \geq 0$. 
\begin{enumerate}
\item {\bf Fidelity} For $\rho_A,\sigma_A \in \mathcal{D}(A)$, $$\F(\rho_A,\sigma_A)\defeq\|\sqrt{\rho_A}\sqrt{\sigma_A}\|_1.$$ For classical probability distributions $P = \{p_i\}, Q =\{q_i\}$, $$\F(P,Q)\defeq \sum_i \sqrt{p_i \cdot q_i}.$$
\item {\bf Purified distance} For $\rho_A,\sigma_A \in \mathcal{D}(A)$, $$\Pur(\rho_A,\sigma_A) = \sqrt{1-\F^2(\rho_A,\sigma_A)}.$$

\item {\bf $\varepsilon$-ball} For $\rho_A\in \mathcal{D}(A)$, $$\ball{\eps}{\rho_A} \defeq \{\rho'_A\in \mathcal{D}(A)|~\Pur(\rho_A,\rho'_A) \leq \varepsilon\}. $$ 
\item {\bf Von-Neumann entropy} For $\rho_A\in\mathcal{D}(A)$, $$S(\rho_A) \defeq - \Tr(\rho_A\log\rho_A) .$$ 
\item {\bf Relative entropy} For $\rho_A,\sigma_A\in \mathcal{D}(A)$ such that $\text{supp}(\rho_A) \subset \text{supp}(\sigma_A)$, $$\relent{\rho_A}{\sigma_A} \defeq \Tr(\rho_A\log\rho_A) - \Tr(\rho_A\log\sigma_A) .$$ 

\item {\bf Relative entropy variance} For $\rho_A,\sigma_A\in \mathcal{D}(A)$ such that $\text{supp}(\rho_A) \subset \text{supp}(\sigma_A)$, $$V(\rho\|\sigma) = \Tr(\rho(\log\rho - \log\sigma)^2) - (\relent{\rho}{\sigma})^2.$$

\item {\bf Max-relative entropy} For $\rho_A,\sigma_A\in \mathcal{P}(A)$ such that $\text{supp}(\rho_A) \subset \text{supp}(\sigma_A)$, $$ \dmax{\rho_A}{\sigma_A}  \defeq  \inf \{ \lambda \in \mathbb{R} : 2^{\lambda} \sigma_A \geq \rho_A \}  .$$  
\item {\bf Smooth max-relative entropy} For $\rho_A\in \mathcal{D}(A) ,\sigma_A\in \mathcal{P}(A)$ such that $\text{supp}(\rho_A) \subset \text{supp}(\sigma_A)$, $$ \dmaxeps{\rho_A}{\sigma_A}{\eps}  \defeq  \sup_{\rho'_A\in \ball{\eps}{\rho_A}} \dmax{\rho_A'}{\sigma_A}  .$$  
\item {\bf Smooth min-relative entropy}  For $\rho_A\in \mathcal{D}(A) ,\sigma_A\in \mathcal{P}(A)$, $$ \dmineps{\rho_A}{\sigma_A}{\eps}  \defeq  \sup_{0<\Pi<I, \Tr(\Pi\rho_A)\geq 1-\eps^2}\log(\frac{1}{\Tr(\Pi\sigma_A)}).$$  

\item {\bf Information spectrum relative entropy} For $\rho_A\in \mathcal{D}(A) ,\sigma_A\in \mathcal{P}(A)$ such that $\text{supp}(\rho_A) \subset \text{supp}(\sigma_A)$, $$ \dseps{\rho_A}{\sigma_A}{\eps}  \defeq  \sup \{R: \Tr(\rho_A\{\rho_A-2^R\sigma_A\}_{+}) \geq 1-\eps \}  .$$  

\item {\bf Information spectrum relative entropy [Alternate definition]} For $\rho_A\in \mathcal{D}(A) ,\sigma_A\in \mathcal{P}(A)$ such that $\text{supp}(\rho_A) \subset \text{supp}(\sigma_A)$, $$ \dsepsalt{\rho_A}{\sigma_A}{\eps}  \defeq  \inf \{R: \Tr(\rho_A\{\rho_A-2^R\sigma_A\}_{-}) \geq 1-\eps \}  .$$

\item {\bf Max-information} For $\rho_{AB}\in \mathcal{D}(AB)$, define 
$$\imax(A:B)_{\rho} = \dmax{\rho_{AB}}{\rho_A\otimes \rho_B}.$$

\item {\bf Smooth max-information} For $\rho_{AB}\in \mathcal{D}(AB)$, define 
$$\imaxepss{\eps}(A:B)_{\rho} = \min_{\rho'\in\ball{\eps}{\rho}}\imax(A:B)_{\rho'} .$$

\item {\bf Smooth max-information [Alternate definition]} For $\rho_{AB}\in \mathcal{D}(AB)$, define 
$$\imaxeps{\eps}(A:B)_{\rho} = \min_{\rho'\in\ball{\eps}{\rho}}\dmax{\rho'_{AB}}{\rho'_A\otimes \rho_B}.$$

\item {\bf Restricted smooth max-information} For $\rho_{AB}\in \mathcal{D}(AB)$, define 
$$\imaxepsbeta{\eps}{\delta}(A:B)_{\rho} = \min_{\rho'\in\ball{\eps}{\rho}: \rho'_A\leq (1+\delta)\rho_A, \rho'_B\leq (1+\delta)\rho_B}\dmax{\rho'_{AB}}{\rho_A\otimes \rho_B}.$$

\suppress{
\item {\bf Mutual information} For $\rho_{AB}\in \mathcal{D}(AB)$, $$\mutinf{A}{B}_{\rho}\defeq S(\rho_A) + S(\rho_B)-S(\rho_{AB}) = \relent{\rho_{AB}}{\rho_A\otimes\rho_B}.$$
\item {\bf Conditional mutual information} For $\rho_{ABC}\in\mathcal{D}(ABC)$, $$\condmutinf{A}{B}{C}_{\rho}\defeq \mutinf{A}{BC}_{\rho}-\mutinf{A}{C}_{\rho}.$$
\item {\bf Max-information}  For $\rho_{AB}\in \mathcal{D}(AB)$, $$ \imax(A:B)_{\rho} \defeq   \inf_{\sigma_{B}\in \mathcal{D}(B)}\dmax{\rho_{AB}}{\rho_{A}\otimes\sigma_{B}} .$$
\item {\bf Smooth max-information} For $\rho_{AB}\in \mathcal{D}(AB)$,  $$\imaxeps(A:B)_{\rho} \defeq \inf_{\rho'\in \ball{\eps}{\rho}} \imax(A:B)_{\rho'} .$$	
}
\end{enumerate}
\label{def:infquant}
\end{definition}	
We will use the following facts. 
\begin{fact}[Triangle inequality for purified distance,~\cite{Tomamichel12}]
\label{fact:trianglepurified}
For states $\rho_A, \sigma_A, \tau_A\in \mathcal{D}(A)$,
$$\Pur(\rho_A,\sigma_A) \leq \Pur(\rho_A,\tau_A)  + \Pur(\tau_A,\sigma_A) . $$ 
\end{fact}
\suppress{
\begin{fact}[\cite{stinespring55}](\textbf{Stinespring representation})\label{stinespring}
Let $\E(\cdot): \mathcal{L}(A)\rightarrow \mathcal{L}(B)$ be a quantum operation. There exists a register $C$ and an unitary $U\in \mathcal{U}(ABC)$ such that $\E(\omega)=\Tr_{A,C}\br{U (\omega  \otimes \ketbra{0}^{B,C}) U^{\dagger}}$. Stinespring representation for a channel is not unique. 
\end{fact}
}
\begin{fact}[Monotonicity under quantum operations, \cite{barnum96},\cite{lindblad75}]
	\label{fact:monotonequantumoperation}
For quantum states $\rho$, $\sigma \in \mathcal{D}(A)$, and quantum operation $\cE(\cdot):\mathcal{L}(A)\rightarrow \mathcal{L}(B)$, it holds that
\begin{align*}
	\dmax{\cE(\rho)}{\cE(\sigma)} \leq \dmax{\rho}{\sigma} \quad \mbox{and} \quad \F(\cE(\rho),\cE(\sigma)) \geq \F(\rho,\sigma) \quad \mbox{and} \quad \dmineps{\rho}{\sigma}{\eps} \geq \dmineps{\cE(\rho)}{\cE(\sigma)}{\eps}.
\end{align*}
In particular, for bipartite states $\rho_{AB},\sigma_{AB}\in \mathcal{D}(AB)$, it holds that
\begin{align*}
	\dmax{\rho_{AB}}{\sigma_{AB}} \geq \dmax{\rho_A}{\sigma_A} \quad \mbox{and} \quad \F(\rho_{AB},\sigma_{AB}) \leq \F(\rho_A,\sigma_A) \quad \mbox{and} \quad \dmineps{\rho_{AB}}{\sigma_{AB}}{\eps} \geq \dmineps{\rho_A}{\sigma_A}{\eps}.
\end{align*}
\end{fact}

\begin{fact}[Uhlmann's Theorem, \cite{uhlmann76}]
\label{uhlmann}
Let $\rho_A,\sigma_A\in \mathcal{D}(A)$. Let $\rho_{AB}\in \mathcal{D}(AB)$ be a purification of $\rho_A$ and $\ket{\sigma}_{AC}\in\mathcal{D}(AC)$ be a purification of $\sigma_A$. There exists an isometry $V: C \rightarrow B$ such that,
 $$\F(\ketbra{\theta}_{AB}, \ketbra{\rho}_{AB}) = \F(\rho_A,\sigma_A) ,$$
 where $\ket{\theta}_{AB} = (\id_A \otimes V) \ket{\sigma}_{AC}$.
\end{fact}

%
\begin{fact}[Pinsker's inequality, \cite{CastelleHR78}]
\label{pinsker}
For quantum states $\rho_A,\sigma_A\in\mathcal{D}(A)$, 
$$\F(\rho,\sigma) \geq 2^{-\frac{1}{2}\relent{\rho}{\sigma}}.$$
\end{fact}

\begin{fact}[Alicki-Fannes inequality, \cite{fannes73}]
\label{fact:fannes}
Given bipartite quantum states $\rho_{AB},\sigma_{AB}\in \mathcal{D}(AB)$, and $\Pur(\rho_{AB},\rho_{AB})= \eps \leq \frac{1}{2\mathrm{e}}$, it holds that $$|\mutinf{A}{B}_{\rho}-\mutinf{A}{B}_{\sigma}|\leq 6\eps\log(|A|)+4.$$   
\end{fact}

\begin{fact}[Triangle property of smooth max- relative entropy]
\label{triangledmax}
For $\rho_A \in \mathcal{D}(A),\sigma_A,\tau_A\in\mathcal{P}(A)$, it holds that
$$\dmaxeps{\rho_A}{\tau_A}{\eps} \leq \dmax{\sigma_A}{\tau_A} + \dmaxeps{\rho_A}{\sigma_A}{\eps}.$$
\end{fact}
\begin{proof}
Let $k\defeq \dmax{\sigma_A}{\tau_A}$, which implies that $\sigma_A \leq 2^k\tau_A$. Let $\rho'_A\in\ball{\eps}{\rho_A}$ be the state achieving the infimum in $R\defeq \dmaxeps{\rho_A}{\sigma_A}{\eps}$. Then $\rho'_A \leq 2^R\sigma_A \leq 2^{R+k}\tau_A$. This implies that $\dmax{\rho'_A}{\tau_A}\leq R+k$, which concludes the fact using the inequality $\dmaxeps{\rho_A}{\tau_A}{\eps} \leq \dmax{\rho'_A}{\tau_A}$.

\end{proof}

\begin{fact}[Gentle measurement lemma,\cite{Winter:1999,Ogawa:2002}]
\label{gentlelemma}
Let $\rho$ be a quantum state and $0<A<\id$ be an operator. Then 
$$\F(\rho, \frac{A\rho A}{\Tr(A^2\rho)})\geq \sqrt{\Tr(A^2\rho)}.$$
\end{fact}
\begin{proof}
Let $\ket{\rho}$ be a purification of $\rho$. Then $(\id \otimes A)\ket{\rho}$ is a purification of $A\rho A$. Now, applying monotonicity of fidelity under quantum operations (Fact \ref{fact:monotonequantumoperation}), we find 
$$\F(\rho, \frac{A\rho A}{\Tr(A^2\rho)}) \geq \F(\ketbra{\rho}, \frac{(\id \otimes A)\ketbra{\rho}(\id\otimes A^{\dagger})}{\Tr(A^2\rho)}) = \sqrt{\frac{\Tr(A\rho)^2}{\Tr(A^2\rho)}}>\sqrt{\Tr(A^2\rho)}.$$
In last inequality, we have used $A>A^2$.
\end{proof}

\begin{fact}[Hayashi-Nagaoka inequality, \cite{HyashiN03}]
\label{haynag}
Let $0<S<\id,T$ be positive semi-definite operators. Then 
$$\id - (S+T)^{-\frac{1}{2}}S(S+T)^{-\frac{1}{2}}\leq 2(\id-S) + 4T.$$

\end{fact}

\begin{fact}[\cite{TomHay13, li2014}]
\label{dmaxequi}
Let $\eps\in (0,1)$ and $n$ be an integer. Let $\rho^{\otimes n}, \sigma^{\otimes n}$ be quantum states. Define $\Phi(x) = \int_{-\infty}^x \frac{e^{-t^2/2}}{\sqrt{2\pi}} dt$. It holds that
\begin{equation*}
\dmaxeps{\rho^{\otimes n}}{\sigma^{\otimes n}}{\eps} = n\relent{\rho}{\sigma} + \sqrt{n\varrelent{\rho}{\sigma}} \Phi^{-1}(\eps) + O(\log n) ,
\end{equation*}
and 
\begin{equation*}
\dmineps{\rho^{\otimes n}}{\sigma^{\otimes n}}{\eps} = n\relent{\rho}{\sigma} + \sqrt{n\varrelent{\rho}{\sigma}} \Phi^{-1}(\eps) + O(\log n) .
\end{equation*}
\end{fact}

\begin{fact}
\label{gaussianupper}
For the function $\Phi(x) = \int_{-\infty}^x \frac{e^{-t^2/2}}{\sqrt{2\pi}} dt$ and $\eps\leq \frac{1}{2}$, it holds that $|\Phi^{-1}(\eps)| \leq 2\sqrt{\log\frac{1}{2\eps}}$.
\end{fact}
\begin{proof}
We have $$\Phi(-x)=\int_{-\infty}^{-x} \frac{e^{-t^2/2}}{\sqrt{2\pi}} dt = \int_{0}^{\infty} \frac{e^{-(-x-t)^2/2}}{\sqrt{2\pi}} dt \leq e^{-x^2/2} \int_{0}^{\infty} \frac{e^{-(-t)^2/2}}{\sqrt{2\pi}} dt = \frac{1}{2}e^{-x^2/2}.$$ Thus, $\Phi^{-1}(\eps) \geq -2\sqrt{\log\frac{1}{2\eps}}$, which completes the proof.
\end{proof}

Following fact says that if a collection of quantum operations do not change a given state much, then successive application of them brings limited change.

\begin{fact}[Fact 21, \cite{AnshuJH17}]
\label{slowchange}
Let $\rho_1$ be a quantum state and $\{\cE_2,\cE_3, \ldots\}$ be a collection of quantum maps. Define a series of quantum states $\{\rho_2,\rho_3,\ldots \}$ recursively as $\rho_i = \cE_i(\rho_{i-1})$. It holds that $$\Pur(\rho_i,\rho_1) \leq (i-1)\max_i\{\Pur(\cE_i(\rho_1),\rho_1)\}.$$
\end{fact}

We shall also need the following series of results, that are central to our achievability approach.

\begin{fact}[~\cite{ADJ14}]
\label{relentconcav}
Let $\mu_1,\mu_2,\ldots \mu_n, \theta$ be quantum states and $\{p_1,p_2,\ldots p_n\}$ be a probability distribution. Let $\mu=\sum_i p_i\mu_i$ be the average state. Then 
$$\relent{\mu}{\theta} =\sum_i p_i(\relent{\mu_i}{\theta}-\relent{\mu_i}{\mu}).$$
\end{fact}

\begin{lemma}
\label{err}
Let $\rho$ and $\sigma$ be quantum states. Then, for every let $0< \Lambda< \mathbb{I}$ be an operator,
\beq
|\sqrt{\tr\left[\Lambda\rho\right]}-\sqrt{\tr\left[\Lambda\sigma\right]}| \leq \Pur(\rho,\sigma). \nonumber
\enq
\end{lemma}
\begin{proof}
Let $\theta, \phi \in [0, \frac{\pi}{2}]$ be such that $\tr\left[\Lambda\rho\right]:= \cos(\theta)$ and $\tr\left[\Lambda\sigma\right]:= \cos(\phi).$ Thus, using monotonicity of fidelity we have
\begin{align}
\F(\rho, \sigma) &= \sqrt{\Tr(\Lambda\rho)\Tr(\Lambda\sigma)} + \sqrt{(1-\Tr(\Lambda\rho))(1-\Tr(\Lambda\sigma))}\geq \F(\rho,\sigma) \geq \sqrt{1-\eps^2}, \nonumber\\
&= \cos(\theta)\cos(\phi) + \sin(\theta)\sin(\theta) \nonumber\\
&= \cos(\theta - \phi) \nonumber\\
&= \sqrt{1- \sin^2(\theta - \phi)} \nonumber\\
\label{pf}
& \leq \sqrt{1 - \left(\cos(\theta) -\cos(\phi)\right)^2},
\end{align}
where the last inequality follows because of the following: 
\begin{align*}
|\sin \left(\theta - \phi\right)| &= 2 \bigg|\sin \left(\frac{\theta-\phi}{2}\right)\bigg| ~~ \bigg |\cos \left(\frac{\theta-\phi}{2}\right)\bigg| \\
& \geq 2 \bigg|\sin \left(\frac{\theta-\phi}{2}\right)\bigg| ~~ \cos \left(\frac{\pi - \theta-\phi}{2}\right) \\
& =  2 \bigg|\sin \left(\frac{\theta-\phi}{2}\right)\bigg| \sin \left(\frac{\theta+\phi}{2}\right)\\
&= \bigg|\cos (\theta) - \cos (\phi)\bigg|,
\end{align*}
where the inequality above follows because $\theta, \phi \in [0, \frac{\pi}{2}].$ The claim of the Lemma now follows from \eqref{pf} and the relation between the purified distance and fidelity between two quantum states. 
\end{proof}
\begin{lemma}[Convex-split lemma,\cite{ADJ14}]
Let $\rho_{PQ}\in\mathcal{D}(PQ)$ and $\sigma_Q\in\mathcal{D}(Q)$ be quantum states such that $\text{supp}(\rho_Q)\subset\text{supp}(\sigma_Q)$.  Let $k \defeq \dmax{\rho_{PQ}}{\rho_P\otimes\sigma_Q}$. Define the following state
\begin{equation*}
\tau_{PQ_1Q_2\ldots Q_n} \defeq  \frac{1}{n}\sum_{j=1}^n \rho_{PQ_j}\otimes\sigma_{Q_1}\otimes \sigma_{Q_2}\ldots\otimes\sigma_{Q_{j-1}}\otimes\sigma_{Q_{j+1}}\ldots\otimes\sigma_{Q_n}
\end{equation*}
on $n+1$ registers $P,Q_1,Q_2,\ldots Q_n$, where $\forall j \in [n]: \rho_{PQ_j} = \rho_{PQ}$ and $\sigma_{Q_j}=\sigma_Q$.  Then for $\delta \in (0,1) $ and $ n= \lceil\frac{2^k}{\delta^2}\rceil$,  $$ \Pur(\tau_{PQ_1Q_2\ldots Q_n},\tau_P \otimes \sigma_{Q_1}\otimes\sigma_{Q_2}\ldots \otimes \sigma_{Q_n}) \leq \delta.$$ 
\end{lemma}
We have the following corollary of above lemma.
\begin{corollary}[Corollary of convex-split lemma]
\label{convexcomb}
For an $\eps>0$. Let $\rho_{PQ}\in\mathcal{D}(PQ)$ and $\sigma_Q\in\mathcal{D}(Q)$ be quantum states such that $\text{supp}(\rho_Q)\subset\text{supp}(\sigma_Q)$.  Let $k \defeq \inf_{\rho'\in \ball{\eps}{\rho_{PQ}}}\dmax{\rho'_{PQ}}{\rho'_P\otimes\sigma_Q}$. Define the following state
\begin{equation*}
\tau_{PQ_1Q_2\ldots Q_n} \defeq  \frac{1}{n}\sum_{j=1}^n \rho_{PQ_j}\otimes\sigma_{Q_1}\otimes \sigma_{Q_2}\ldots\otimes\sigma_{Q_{j-1}}\otimes\sigma_{Q_{j+1}}\ldots\otimes\sigma_{Q_n}
\end{equation*}
on $n+1$ registers $P,Q_1,Q_2,\ldots Q_n$, where $\forall j \in [n]: \rho_{PQ_j} = \rho_{PQ}$ and $\sigma_{Q_j}=\sigma_Q$.  
 For $\delta \in (0,1) $ and $ n= \lceil\frac{2^k}{\delta^2}\rceil$, 
 $$ \Pur(\tau_{PQ_1Q_2\ldots Q_n},\tau_P \otimes \sigma_{Q_1}\otimes\sigma_{Q_2}\ldots \otimes \sigma_{Q_n}) \leq 2\eps+\delta.$$ 
\end{corollary}
\begin{proof}
Let $\rho'_{PQ}$ be the state achieving infimum in 
$$ \inf_{\rho'\in \ball{\eps}{\rho_{PQ}}}\dmax{\rho'_{PQ}}{\rho'_P\otimes\sigma_Q}.$$ It holds that $\Pur(\rho_{PQ},\rho'_{PQ})\leq \eps.$
Define the state
\begin{equation*}
\tau'_{PQ_1Q_2\ldots Q_n} \defeq  \frac{1}{n}\sum_{j=1}^n \rho'_{PQ_j}\otimes\sigma_{Q_1}\otimes \sigma_{Q_2}\ldots\otimes\sigma_{Q_{j-1}}\otimes\sigma_{Q_{j+1}}\ldots\otimes\sigma_{Q_n} .
\end{equation*}
Then by convex-split lemma, and the choice of $n$, it holds that 
 $$ \Pur(\tau'_{PQ_1Q_2\ldots Q_n},\tau'_P \otimes \sigma_{Q_1}\otimes\sigma_{Q_2}\ldots \otimes \sigma_{Q_n}) \leq \delta.$$ 
Moreover, using the concavity of fidelity (Theorem $9.7$, \cite{NielsenC00} ), $\Pur(\tau'_{PQ_1Q_2\ldots Q_n}, \tau_{PQ_1Q_2\ldots Q_n}) \leq \Pur(\rho'_{PQ},\rho_{PQ}) \leq \eps$. Similarly, $\Pur(\tau'_P,\tau_P) = \Pur(\tau'_P,\rho_P)\leq \eps$. Thus, by triangle inequality for purified distance (Fact \ref{fact:trianglepurified}), 
$$\Pur(\tau_{PQ_1Q_2\ldots Q_n}, \rho_P \otimes \sigma_{Q_1}\otimes\sigma_{Q_2}\ldots \otimes \sigma_{Q_n}) \leq 2\eps+\delta.$$
\end{proof}

We will use the following new version of the convex split lemma. 
\begin{lemma}[Bi-partite convex-split lemma]
\label{2dconvexcomb}
Let $\rho_{PQ}\in\mathcal{D}(PQ)$ be a quantum state, $\eps,\delta > 0$ and $k \defeq \imaxepsbeta{\eps}{\delta} (P:Q)_{\rho}$. Choose integers $n,m>\frac{1}{\delta}$  and define the following state
\begin{equation*}
\tau_{P_1\ldots P_mQ_1\ldots Q_n} \defeq  \frac{1}{n\cdot m}\sum_{i=1}^n\sum_{j=1}^{m} \rho_{P_iQ_j}\otimes\rho_{P_1}\otimes\ldots\rho_{P_{i-1}}\otimes \rho_{P_{i+1}}\otimes\ldots\rho_{P_m}\otimes\rho_{Q_1}\otimes \ldots\rho_{Q_{j-1}}\otimes\rho_{Q_{j+1}}\ldots\otimes\rho_{Q_n}
\end{equation*}
on registers $P_1,P_2\ldots P_m,Q_1,Q_2,\ldots Q_n$, where $\forall i,j \in [n]: \rho_{P_iQ_j} = \rho_{PQ}$, $\rho_{P_i}=\rho_P$ and $\rho_{Q_j}=\rho_Q$.  Then,  $$ \Pur(\tau_{P_1P_2\ldots P_mQ_1Q_2\ldots Q_n},\rho_{P_1}\otimes\rho_{P_2}\otimes\ldots\rho_{P_m} \otimes \rho_{Q_1}\otimes\rho_{Q_2}\ldots \otimes \rho_{Q_n}) \leq \eps+2\sqrt{\delta}+\sqrt{\frac{2^k}{n\cdot m}}.$$ In particular, if it is possible to further choose $n,m$ such that $ n\cdot m\geq \lceil\frac{2^k}{\delta}\rceil$, we find that
  $$ \Pur(\tau_{P_1P_2\ldots P_mQ_1Q_2\ldots Q_n},\rho_{P_1}\otimes\rho_{P_2}\otimes\ldots\rho_{P_m} \otimes \rho_{Q_1}\otimes\rho_{Q_2}\ldots \otimes \rho_{Q_n}) \leq \eps+3\sqrt{\delta}.$$ 
\end{lemma}

The proof closely follows the original proof of convex split lemma from~\cite{ADJ14}.
\begin{proof}

Let $\rho'_{PQ}$ be the quantum state achieving the optimum in the definition of $\imaxepsbeta{\eps}{\delta} (P:Q)_{\rho}$. We shall work with the state 
\begin{equation*}
\tau'_{P_1\ldots P_mQ_1\ldots Q_n} \defeq  \frac{1}{n\cdot m}\sum_{i=1}^n\sum_{j=1}^{m} \rho'_{P_iQ_j}\otimes\rho_{P_1}\otimes\ldots\rho_{P_{i-1}}\otimes \rho_{P_{i+1}}\otimes\ldots\rho_{P_m}\otimes\rho_{Q_1}\otimes \ldots\rho_{Q_{j-1}}\otimes\rho_{Q_{j+1}}\ldots\otimes\rho_{Q_n}
\end{equation*}

Define, 
$$\rho^{-(i,j)} \defeq \rho_{P_1}\otimes\ldots\rho_{P_{i-1}}\otimes \rho_{P_{i+1}}\otimes\ldots\rho_{P_m}\otimes\rho_{Q_1}\ldots\otimes\rho_{Q_{j-1}}\otimes \rho_{Q_{j+1}}\ldots \otimes \rho_{Q_n},$$ 
$$\rho\defeq \rho_{P_1}\otimes\rho_{P_2}\otimes\ldots\rho_{P_m}\otimes  \rho_{Q_1}\otimes \rho_{Q_2}\ldots \rho_{Q_n}. $$ Then 
$$\tau'_{P_1P_2\ldots P_mQ_1Q_2\ldots Q_n}=\frac{1}{n\cdot m}\sum_{i,j}\rho'_{P_iQ_j}\otimes \rho^{-(i,j)}. $$ 
Now, we from Fact \ref{relentconcav} we have the following:
\begin{align}
& \relent{\tau'_{P_1\ldots P_mQ_1\ldots Q_n}}{\rho} \nonumber\\
\label{eq:convsplit}
& = \frac{1}{n\cdot m}\sum_{i,j} \relent{\rho'_{P_iQ_j}\otimes \rho^{-(i,j)}}{\rho}  -\frac{1}{n\cdot m}\sum_{i,j}\relent{\rho'_{P_iQ_j}\otimes \rho^{-(i,j)}}{\tau'_{P_1P_2\ldots P_mQ_1Q_2\ldots Q_n}}.
\end{align}
Note that,
$$\relent{\rho'_{P_iQ_j}\otimes \rho^{-(i,j)}}{\rho} = \relent{\rho'_{P_iQ_j}}{\rho_{P_i}\otimes \rho_{Q_j}}  \mbox{ and } \relent{\rho'_{P_iQ_j}\otimes \rho^{-(i,j)}}{\tau'_{P_1P_2\ldots P_mQ_1Q_2\ldots Q_n}} \geq \relent{\rho'_{P_iQ_j}}{\tau'_{P_iQ_j}},$$ 
as relative entropy decreases under partial trace. Further,
$$\tau'_{P_iQ_j} = \frac{1}{n\cdot m}\rho'_{P_iQ_j}+ \frac{1}{n}(1-\frac{1}{m})\rho'_{P_i}\otimes \rho_{Q_j} + \frac{1}{m}(1-\frac{1}{n})\rho_{P_i}\otimes \rho'_{Q_j}+ (1-\frac{1}{n}-\frac{1}{m}+\frac{1}{n\cdot m})\rho_{P_i}\otimes\rho_{Q_j}.$$ By assumption, $\rho'_{P_iQ_j} \leq 2^k\rho_{P_i}\otimes \rho_{Q_j}, \rho'_{P_i}\leq (1+\delta)\rho_{P_i}, \rho'_{Q_i}\leq (1+\delta)\rho_{Q_i}$. Hence $\tau'_{P_iQ_j} \leq (1+ \frac{1+\delta}{n}+\frac{1+\delta}{m}+\frac{2^k-1}{n\cdot m})\rho_{P_i}\otimes\rho_{Q_j}$. Since $\log(\cdot)$ is operator monotone, we have
\begin{align}
&\relent{\rho_{P_iQ_j}}{\tau'_{P_iQ_j}} \nonumber\\
& = \Tr(\rho_{P_iQ_j}\log\rho_{P_iQ_j}) - \Tr(\rho_{P_iQ_j}\log\tau'_{P_iQ_j}) \nonumber\\ 
&\geq \Tr(\rho_{P_iQ_j}\log\rho_{P_iQ_j})  - \Tr(\rho_{P_iQ_j}\log(\rho_{P_i}\otimes\rho_{Q_j})) - \log\left(1+ \frac{1+\delta}{n}+\frac{1+\delta}{m}+\frac{2^k-1}{n\cdot m}\right) \nonumber\\ 
\label{cb1}
& = \relent{\rho_{P_iQ_j}}{\rho_{P_i}\otimes\rho_{Q_j}}- \log\left(1 + \frac{1+\delta}{n}+\frac{1+\delta}{m}+ \frac{2^k-1}{n\cdot m}\right) .
\end{align}
We now have the following upper bound on Equation \ref{eq:convsplit}:
\begin{align*}
&\relent{\tau'_{P_1P_2\ldots P_mQ_1Q_2\ldots Q_n}}{\rho}\\
&\leq \frac{1}{n\cdot m}\sum_{i,j} \relent{\rho_{P_iQ_j}}{\rho_{P_i}\otimes\rho_{Q_j}} - \frac{1}{n\cdot m}\sum_{i,j}\relent{\rho_{P_iQ_j}}{\rho_{P_i}\otimes\rho_{Q_j}}+ \log\left(1 + \frac{1+\delta}{n}+\frac{1+\delta}{m}+ \frac{2^k-1}{n\cdot m}\right)\\
& \leq \log\left(1 + 2\cdot\delta(1+\delta)+ \frac{2^k-1}{n\cdot m}\right),
\end{align*}
where the first inequality above follows from \eqref{cb1}.
Above, the last inequality follows by the choice of $n,m$. Thus, by Pinsker's inequality (Fact \ref{pinsker}), we obtain that 
$$\Pur(\tau'_{P_1P_2\ldots P_mQ_1Q_2\ldots Q_n},\rho) \leq \sqrt{2\cdot\delta(1+\delta)+ \frac{2^k}{n\cdot m}} \leq 2\sqrt{\delta} + \sqrt{\frac{2^k}{n\cdot m}}.$$

Since $\Pur(\tau'_{P_1P_2\ldots P_mQ_1Q_2\ldots Q_n},\tau_{P_1P_2\ldots P_mQ_1Q_2\ldots Q_n})\leq \Pur(\rho'_{PQ},\rho_{PQ})\leq \eps$, triangle inequality for purified distance (Fact \ref{fact:trianglepurified}) shows that 

$$\Pur(\tau'_{P_1P_2\ldots P_mQ_1Q_2\ldots Q_n},\rho) \leq \eps+2\sqrt{\delta}+\sqrt{\frac{2^k}{n\cdot m}}.$$

This proves first part of the lemma. The second part follows from our choice of $n\cdot m$.

\end{proof}

\section{Point to point channel}
\label{sec:ptop}
\subsection*{Description of task} 

\begin{figure}[h]
\centering
\includegraphics[scale=0.4]{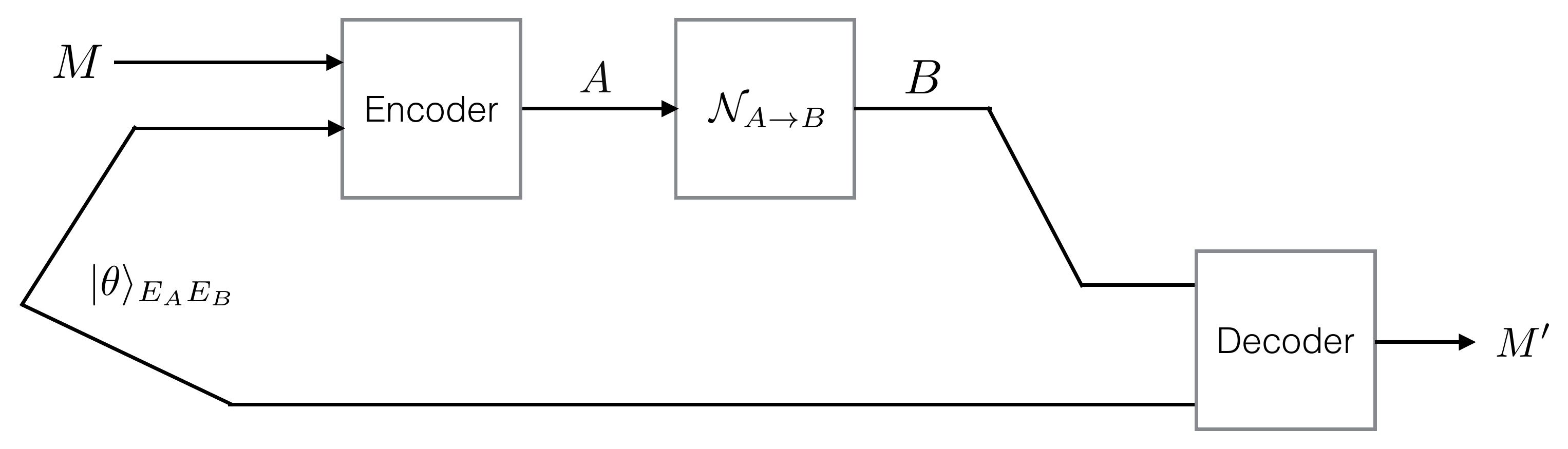}
\caption{A sketch of general entanglement assisted protocol for point to point channel}
\label{fig:pointtopoint}
\end{figure}

There are two parties $\alice$  and $\bob$ . $\alice$  wants to communicate a classical message $M$ chosen from $[1:2^R]$ to $\bob$  over a quantum channel such that $\bob$  is able to decode the correct message with probability at least $1-\eps^2$ , for all message $m$. To accomplish this task $\alice$  and $\bob$  also share entanglement between them.  Let the input to $\alice$  be given in a register $M$. We now make the following definition, illustrated in Figure~\ref{fig:pointtopoint}:
\begin{definition}
\label{codeptop}
Let $\ket{\theta}_{E_AE_B}$ be the shared entanglement between $\alice$  and $\bob$ . An $(R, \eps )$-entanglement assisted code for the quantum channel $\cN_{ A \to B}$ consists of 
\begin{itemize}
\item An encoding unitary $\cE: ME_A \rightarrow A $ for $\alice$ .  
\item A decoding operation $\cD : B E_B\rightarrow M'$ for $\bob$, with $M'\equiv M$ being the output register such that for all $m$,
\beq
\Pr(M'\neq m|M=m) \leq \eps^2. \nonumber
\enq
\end{itemize}
\end{definition}

\subsection*{An achievability protocol}

We show the following result. 
\begin{theorem}[Restatement of Theorem \ref{theo:ptopintro}]
\label{theo:achieveptop}
Let $\cN_{A \to B }$ be the quantum channel and let $\eps,\delta \in (0,1)$. Let $A'\equiv A$ be a purifying register. Then, for any $R$ satisfying 
\beq
\label{eq:pointchannelrate}
R \leq  \max _{\ketbra{\psi}_{A A'}}\dmineps{\cN_{A \to B } (\ketbra{\psi}_{AA'})}{\cN_{A\to B}(\psi_{A}) \otimes \psi_{A'}}{\eps}- 2\log \frac{1}{\delta},
\enq
there exists an $(R,2\eps + 2\delta)$ entanglement assisted code for the quantum channel $\cN_{A \to B }.$
\end{theorem}

\begin{figure}[h]
\centering
\includegraphics[scale=0.4]{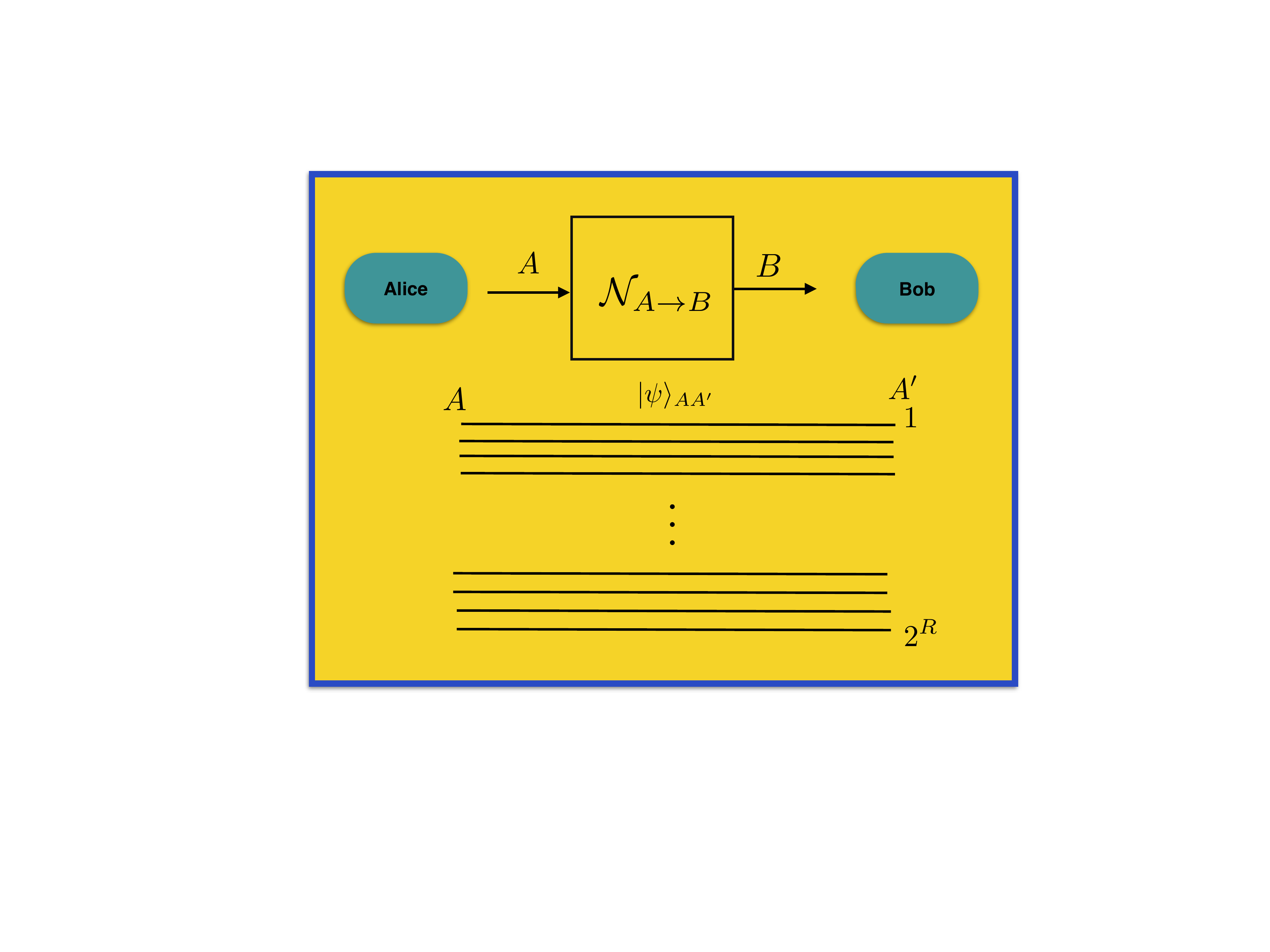}
\caption{A schematic for the achievability protocol. Upon receiving the message $m$, $\alice$ sends the $m$-th register in the entanglement through the channel to $\bob$.}
\label{fig:pointtopointprotocol}
\end{figure}

\begin{proof}
Fix $\psi_{AA'}$ and $R$ as given in Equation \ref{eq:pointchannelrate}. Introduce the registers $A_1,A_2,\ldots A_{2^R}$, such that $A_i\equiv A$ and $A'_1,A'_2,\ldots A'_{2^R}$ such that $A'_i\equiv A'$.  $\alice$  and $\bob$  share the state $$\ketbra{\psi}_{A_1A'_1}\otimes \ketbra{\psi}_{A_2A'_2},\ldots \ketbra{\psi}_{A_{2^R}A'_{2^R}},$$
 where $\alice$  holds the registers $A_1,A_2, \cdots, A_{2^{R}}$ and $\bob$  holds the registers $A'_1,A'_2, \cdots, A'_{2^{R}}$. Let, $ 0 \preceq \Pi_{BA'} \preceq \mathrm{I}$ be such that 
\beq
\label{optimalmeasurement}
\dmineps{\cN_{A \to B } (\ketbra{\psi}_{AA'})}{\cN_{A\to B}(\psi_{A}) \otimes \psi_{A'}}{\eps}:= - \log \tr \left[\Pi_{BA'} \cN_{A\to B}(\psi_{A}) \otimes \psi_{A'} \right ]. \nonumber
\enq
Our protocol is as follows (see also Figure~\ref{fig:pointtopointprotocol}): 

\vspace{2mm}

\noindent {\bf{Encoding:}} $\alice$  on receiving the message $m \in [1:2^R]$ sends the register $A_m$ over the channel. Notice that after this transmission over the channel the state in $\bob$ 's possession is the following:
\begin{equation*}
\hat{{\Theta}}_{ B,A'_1\cdots,A'_{2^{R}}}:= \psi_{A'_1}\otimes  \cdots\otimes \cN^{ A_m \to B}\left( \psi_{A_mA'_m}\right)\cdots \otimes \psi_{A'_{2^R}}.
\end{equation*}
Further, notice that $\hat{{\Theta}}_{BA^\prime_j}$ state between the register $A'_j$ and the channel output $B$ is the following  
\begin{equation*}
\label{jointstate}
\hat{{\Theta}}_{BA^\prime_j}   =
\begin{cases}
\cN_{A \to B} (\ketbra{\psi}_{AA'}) & \mbox{if }~ j =m; \\
\cN_{A \to B}(\psi_{A}) \otimes \psi_{A'}        & \mbox{otherwise.}
\end{cases}
\end{equation*}
 {\bf{Decoding:}} For each $m \in [1:2^{nR}],$ we have the operators $\Pi_{A'_m}$ as defined in \eqref{optimalmeasurement}. Using this, we define for each $m$, 
\begin{equation*}
\Lambda(m):= \mathrm{I}_{A'_1} \otimes \mathrm{I}_{A'_2} \otimes \cdots \Pi_{BA'_m} \otimes \cdots \otimes \mathrm{I}_{A'_{2^R}}
\end{equation*}
The decoding POVM element corresponding to $m $ is:
 
\begin{equation*}
\Omega(m) := \left(\sum_{ m^\prime \in [1:2^{R}]} \Lambda({m'})\right)^{-\frac{1}{2}}\Lambda({m})\left(\sum_{ m^\prime \in [1:2^{R}]} \Lambda({m'})\right)^{-\frac{1}{2}}. 
\end{equation*}
It is easy to observe that $\sum_m \Omega(m) = \mathrm{I}$, and hence it forms a valid POVM.

\vspace{2mm}

\noindent {\bf{Probability of error:}} Let $M$ be the message which was transmitted by $\alice$  using the strategy above and let $M'$ be the decoded message by $\bob$  using the above mentioned decoding POVMs. Notice that by the symmetry of the encoding and decoding strategy, it is enough to show that $\Pr \left\{M' \neq 1 \mid  M=1 \right\} \leq 2\eps + 4\delta$, under the event that $M=1$ is the transmitted message. 
\begin{align*}
\Pr \left\{M' \neq 1 | M=1\right\}& = \tr \left[\left(\mathrm{I} -\Omega(1)\right)\hat{{\Theta}}_{ A^\prime_1,B_1\cdots,B_{2^{R}}}\right]\\
& \overset{a} \leq   2\tr \left[\left(\mathrm{I} -\Lambda(1)\right)\hat{{\Theta}}_{ A^\prime_1,B_1\cdots,B_{2^{R}}}\right] + 4\sum_{m\neq 1} \Tr \left[\Lambda(m)\hat{{\Theta}}_{ A^\prime_1,B_1\cdots,B_{2^{R}}} \right]\\
&\overset{b} \leq 2 \eps^2 + 4\times 2^{R - \dmineps{\cN_{A \to B } (\ketbra{\psi}_{AA'})}{\cN_{A\to B}(\psi_{A}) \otimes \psi_{A'}}{\eps}} \\
& \overset{c}\leq 2\eps^2 + 4\delta^2 \leq 4(\eps+\delta)^2 .
\end{align*}
where $a$ follows from Hayashi-Nagaoka operator inequality (Fact \ref{haynag}); $b$ follows from the definition of $\Lambda (m)$, and from the definition of $\Pi_{BA^\prime}$ and $c$ follows from the choice of $R$ as mentioned in the Theorem. This completes the proof. 
\end{proof}

For the same task without shared entanglement, we run the same protocol in which the state $\psi_{AA'}$ is classical-quantum with $A'$ being classical, which is held by $\bob$. We can fix the classical part and obtain a protocol without shared entanglement, for average bounded error under a given input distribution.

\section{Asymptotic and i.i.d. case and its resource requirement}
\label{sec:ptopasymptoticiid}

In the asymptotic and i.i.d. case, we are allowed many uses of the channel. A naive application of Theorem \ref{theo:achieveptop} shows that the amount of resource required is exponentially large in the number of channel use. Here, we shall develop an appropriate asymptotic and i.i.d. version of the position-based decoding and use it to construct an entanglement assisted code with its resource requirement matching that of \cite{BennettSST02}. To measure the performance of this code, we define the rate of communication and the rate of entanglement required in a protocol.
\begin{definition}
\label{codeptopiid}
A pair $(R,E)$ is an achievable pair of communication rate and entanglement rate for the quantum channel $\cN_{A \to B}$, if for every $\eps, \delta \in (0,1)$, there exists a large enough $n$ such that there exists an $(n(R - \delta), \eps)$-entanglement assisted code for the quantum channel $\cN_{A \to B}^{\otimes n}$ with the number of qubits of pre-shared entanglement required at most $n(E+ \delta)$.   
\end{definition}

We will need some notations for our analysis. Let $\cS_{n,w}$ be the set of all subsets of $\{1,2,\ldots w\}$ of size $n$. For an $s\in \cS_{n,w}$, let $s(i)$ be the $i$-th element of $s$, when $s$ is in ascending order. Further, let $\bar{s}$ be the complement of $s$, which is a subset of size $w-n$. Let $\bar{s}(i)$ represent the $i$-th element of $\bar{s}$, when $\bar{s}$ is in ascending order.

The following theorem shall be used later to construct a protocol for entanglement assisted communication over a quantum channel $\cN_{A\to B}$ in the asymptotic and i.i.d. setting, with the pair $(R,E)$ in Definition \ref{codeptopiid} matching with that of \cite{BennettSST02}. It is partly inspired by the work \cite{StrelchukHO13}.

\begin{theorem}
\label{theo:ptopiidachieve}
Let $\cN_{A \to B }$ be the quantum channel, $\eps \in (0,\frac{1}{7})$ and positive integers $n, w$ such that $w> 2n$. Let $A'\equiv A$ be a purifying register. Then, for any quantum state $\psi_{AA'}$ and any $R$ satisfying 
\begin{equation}
\label{eq:iidchannelrate}
nR \leq  \dmineps{\cN_{A \to B }(\psi_{AA'})^{\otimes n}}{\cN_{A\to B}(\psi_{A})^{\otimes n} \otimes \psi^{\otimes n}_{A'}}{\eps}- \frac{n^2\cdot 2^{\dmax{\psi_{AA'}}{\psi_A\otimes \psi_{A'}}}\log e}{w} +\log \eps,
\end{equation}
there exists a $(nR, 7\eps)$-entanglement assisted code for the quantum channel $\cN^{\otimes n}_{A \to B }$. The protocol requires $w$ copies of $\psi_{AA'}$ as shared entanglement. The number of bits of shared randomness used in the protocol is at most $nR + n\log w$.
\end{theorem}
\begin{proof}
Let $\psi_{A A'}$ be the quantum state as given in the statement. Define $F := \dmax{\psi_{AA'}}{\psi_A\otimes \psi_{A'}}$.  Let $\Pi_{B^nA'^n}$ (where $B^n, A'^n$ are $n$ copies of $B, A'$ respectively) be defined as
\begin{equation}
\label{eq:iidPidef}
\Pi_{B^nA'^n} := \underset{\Pi}{\arg \max}\left(\dmineps{\cN_{A \to B }(\psi_{AA'})^{\otimes n}}{\cN_{A\to B}(\psi_{A})^{\otimes n} \otimes \psi^{\otimes n}_{A'}}{\eps}\right).
\end{equation}

Alice and Bob share $w$ copies of $\psi_{A A'}$ in registers $A_1A'_1, A_2A'_2, \ldots A_wA'_w$. Additionally, Alice and Bob share the following randomness in registers $S_1,S_2, \ldots S_{2^{nR}}$, 
$$\sum_{s_1, \ldots s_{2^{nR}}}\barq(s_1,\ldots s_{2^{nR}}) \ketbra{s_1, \ldots s_{2^{nR}}}_{S_1, \ldots S_{2^{nR}}},$$
where $|S_i| = {w \choose n}$, $s_i\in \cS_{n,w}$ and $\barq$ is a pairwise independent probability distribution (that is, $\barq(s_i,s_j) = \barq(s_i)\barq(s_j)$) satisfying $\barq(s_i) = \frac{1}{{w \choose n}}$ for all $s_i$.

\vspace{0.1in}

\noindent {\bf Encoding:} Alice takes a sample from the shared randomness. Let the sample be $s_1, s_2, \ldots s_{2^{nR}}$. To send the message $m \in \{1,2,\ldots 2^{nR}\}$, Alice looks at the set $s_m$ and sends the registers $A_{s_m(1)}, A_{s_m(2)}, \ldots A_{s_m(n)}$ with $n$ uses of the channel.

Further, if Alice sends the message $m$, the quantum state between Bob's registers and the channel output is $$\hat{\Theta}(m) := \cN_{A_{s_m(1)}\to B_1}(\psi_{A_{s_m(1)}A'_{s_m(1)}})\otimes \ldots \cN_{A_{s_m(n)}\to B_n}(\psi_{A_{s_m(n)}A'_{s_m(n)}})\otimes \id_{A'_{\bar{s}(1)}}\otimes \ldots \id_{A'_{\bar{s}(w-n)}},$$ where $\bar{s}$ is the complement of the set $s$.

\vspace{0.1in}

\noindent {\bf Decoding:} Bob takes a sample from the shared randomness. This sample is the same as that obtained by Alice, that is, $s_1, s_2, \ldots s_{2^{nR}}$. For each $m\in \{1,2,\ldots 2^{nR}\}$, define the operator 
$$\Lambda(m) := \Pi_{B_1A'_{s_m(1)}\ldots B_nA'_{s_m(n)}}\otimes \id_{A'_{\bar{s}(1)}}\otimes \ldots \id_{A'_{\bar{s}(w-n)}}.$$  The decoding POVM for message $m$ is 
$$\Omega(m) := \left(\sum_{ m' \in \{1,2,\ldots 2^{nR}\}} \Lambda({m'})\right)^{-\frac{1}{2}}\Lambda({m})\left(\sum_{m' \in \{1,2,\ldots 2^{nR}\}} \Lambda({m'})\right)^{-\frac{1}{2}}.$$ 
It is easy to observe that $\sum_m \Omega(m) \preceq \id$, and hence it forms a valid POVM once the POVM element $\id - \sum_m \Omega(m)$ (interpreted as `no outcome') is added.

\vspace{0.1in}

\noindent{\bf Probability of error:} Let $M$ be the message which was transmitted by Alice  using the strategy above and let $M'$ be the decoded message by Bob  using the above mentioned decoding POVMs. We proceed in a manner similar to the proof of Theorem \ref{theo:achieveptop}. Consider,
\begin{eqnarray}
\label{eq:errorprobcalciid}
&& \Pr \left\{M' \neq m | M=m\right\}= \sum_{s_1,\ldots s_{2^{nR}}}\barq(s_1,\ldots s_{2^{nR}})\tr \left(\left(\mathrm{I} -\Omega(m)\right)\hat{\Theta}(m)\right)\nonumber\\
&&\overset{(1)}\leq \sum_{s_1,\ldots s_{2^{nR}}}\barq(s_1,\ldots s_{2^{nR}})\left( 2\tr \left(\left(\mathrm{I} -\Lambda(m)\right)\hat{\Theta}(m)\right) + 4\sum_{m'\neq m} \Tr \left(\Lambda(m')\hat{\Theta}(m) \right)\right)\nonumber\\
&& \leq 2\eps + 4\sum_{s_1,\ldots s_{2^{nR}}}\barq(s_1,\ldots s_{2^{nR}}) \sum_{m'\neq m} \Tr \left(\Lambda(m')\hat{\Theta}(m) \right)\nonumber\\
&& = 2\eps + 4\sum_{m'\neq m} \sum_{s_m, s_{m'}}\barq(s_m, s_{m'}) \Tr \left(\Lambda(m')\hat{\Theta}(m) \right) \nonumber\\
&& \overset{(2)}= 2\eps + \frac{4}{{w\choose n}^2}\sum_{m'\neq m} \sum_{s_m, s_{m'}} \Tr \left(\Lambda(m')\hat{\Theta}(m) \right)\nonumber\\ && = 2\eps + \frac{4}{{w\choose n}^2}\sum_{m'\neq m} \sum_{s_m, s_{m'}} \Tr \left(\Pi_{B_1A'_{s_{m'}(1)}\ldots B_nA'_{s_{m'}(n)}}\Tr_{A'_{\bar{s}_{m'}(1)}\ldots A'_{\bar{s}_{m'}(w-n)}}\left(\hat{\Theta}(m)\right)\right),\nonumber\\
\end{eqnarray}
where $(1)$ uses the Hayashi-Nagaoka operator inequality (Fact \ref{haynag}) and $(2)$ uses the pairwise independence of $\barq$. Recalling that $$\psi_{AA'}\preceq 2^F\cdot \psi_A\otimes \psi_{A'} \implies \cN_{A\to B}(\psi_{AA'})\preceq 2^F\cdot \cN_{A\to B}(\psi_A)\otimes \psi_{A'},$$ we can substitute
$$\Tr_{A'_{\bar{s}_{m'}(1)}\ldots A'_{\bar{s}_{m'}(w-n)}}\left(\hat{\Theta}(m)\right) \preceq 2^{F\cdot |s_m \cap s_{m'}|} \cdot \cN_{A\to B}(\psi_A)^{\otimes n} \otimes \psi_{A'_{s_{m'}(1)}} \otimes \ldots \psi_{A'_{s_{m'}(n)}}$$ in Equation \ref{eq:errorprobcalciid} to obtain
\begin{eqnarray*}
&& \Pr \left\{M' \neq m | M=m\right\} \leq 2\eps +\frac{4}{{w\choose n}^2} \cdot \\ 
&& \sum_{m'\neq m}\sum_{s_m, s_{m'}} 2^{F\cdot |s_m \cap s_{m'}|}\Tr \left(\Pi_{B_1A'_{s_{m'}(1)}\ldots B_nA'_{s_{m'}(n)}}\cN_{A\to B}(\psi_A)^{\otimes n} \otimes \psi_{A'_{s_{m'}(1)}} \otimes \ldots \psi_{A'_{s_{m'}(n)}}\right) \\
&& \overset{(1)}\leq 2\eps +\frac{4}{{w\choose n}^2} \cdot\sum_{m'\neq m}\sum_{s_m, s_{m'}} 2^{F\cdot|s_m \cap s_{m'}|}\cdot 2^{-\dmineps{\cN_{A \to B }(\ketbra{\psi}_{AA'})^{\otimes n}}{\cN_{A\to B}(\psi_{A})^{\otimes n} \otimes \psi^{\otimes n}_{A'}}{\eps}}\\
&& \leq 2\eps +\frac{4}{{w\choose n}^2} \cdot\sum_{s_m, s_{m'}} 2^{F\cdot|s_m \cap s_{m'}|}\cdot 2^{nR-\dmineps{\cN_{A \to B }(\ketbra{\psi}_{AA'})^{\otimes n}}{\cN_{A\to B}(\psi_{A})^{\otimes n} \otimes \psi^{\otimes n}_{A'}}{\eps}}\\
&&= 2\eps + \left(\frac{4}{{w \choose n}}\sum_{t=0}^n {n \choose t}\cdot {w-n \choose n-t} 2^{F\cdot t}\right)\cdot 2^{nR-\dmineps{\cN_{A \to B }(\ketbra{\psi}_{AA'})^{\otimes n}}{\cN_{A\to B}(\psi_{A})^{\otimes n} \otimes \psi^{\otimes n}_{A'}}{\eps}}\\
&&= 2\eps + 4\cdot\left(\frac{w-n!^2}{w! w-2n!}\sum_{t=0}^n {n \choose t}\frac{w-2n!}{w-2n+t!}\frac{n!}{n-t!} 2^{F\cdot t}\right)\cdot \\ && \hspace{5mm} 2^{nR-\dmineps{\cN_{A \to B }(\ketbra{\psi}_{AA'})^{\otimes n}}{\cN_{A\to B}(\psi_{A})^{\otimes n} \otimes \psi^{\otimes n}_{A'}}{\eps}}\\
&&\leq 2\eps + 4\cdot\left(\sum_{t=0}^n {n \choose t}\left(\frac{n}{w-2n}\right)^t 2^{F\cdot t}\right)\cdot 2^{nR-\dmineps{\cN_{A \to B }(\ketbra{\psi}_{AA'})^{\otimes n}}{\cN_{A\to B}(\psi_{A})^{\otimes n} \otimes \psi^{\otimes n}_{A'}}{\eps}}\\
&&= 2\eps + 4\cdot (1+\frac{2^Fn}{w-2n})^n\cdot 2^{nR-\dmineps{\cN_{A \to B }(\ketbra{\psi}_{AA'})^{\otimes n}}{\cN_{A\to B}(\psi_{A})^{\otimes n} \otimes \psi^{\otimes n}_{A'}}{\eps}}\\
&& \leq 2\eps + 4\cdot 2^{\frac{(2^F\log e)\cdot n^2}{w-2n} + nR- \dmineps{\cN_{A \to B }(\ketbra{\psi}_{AA'})^{\otimes n}}{\cN_{A\to B}(\psi_{A})^{\otimes n} \otimes \psi^{\otimes n}_{A'}}{\eps}}. 
\end{eqnarray*}
Here, $(1)$ uses Equation \ref{eq:iidPidef} and rest of the equations follow by simple analysis. Last inequality follows from the relation $1+x \leq 2^{x\log e}$.  The desired upper bound now follows by the definition of $R$. The upper bound on the number of bits of shared randomness follows from the explicit construction of pairwise independent random variables given in \cite[Section 3]{Lovettnotes}. 
\end{proof}

We observe that in above protocol, only $n$ copies of the shared entanglement (that is, $\ketbra{\psi}_{AA'}$) are used and rest of the shared copies are close to the original with fidelity $1-\eps$ . This shows that rest of the shared copies serve as \textit{catalysts}. In fact, this observation allows us to prove the following improved result.

\begin{theorem}
\label{theo:ptopiidachieve2}
Let $n, b$ be positive integers such that $b<n$ and $\eps, \delta \in (0,1)$. For every quantum state $\psi_{AA'}$, there exists an $(nR, \eps)$ entanglement assisted code for the channel $\cN_{A\to B}^{\otimes n}$, if $R$ satisfies 
\begin{eqnarray*}
R &\leq& \relent{\cN_{A \to B }(\psi_{AA'})}{\cN_{A\to B}(\psi_{A})\otimes \psi_{A'}}- \delta - \frac{b}{n}O(\log \frac{n b^2}{\eps}) \\ &&- \sqrt{\frac{4\log\frac{b^4}{\eps}}{n}\varrelent{\cN_{A \to B }(\psi_{AA'})}{\cN_{A\to B}(\psi_{A})\otimes \psi_{A'}}}. 
\end{eqnarray*} 
The code requires $w$ copies of the shared entanglement $\psi_{AA'}$, where $w$ satisfies 
$$w = n + \frac{3n\cdot 2^{\dmax{\psi_{AA'}}{\psi_A\otimes \psi_{A'}}}}{b\delta}.$$ The number of bits of shared randomness required is equal to $nR + n\log w$.
\end{theorem}

\begin{proof}[Proof of Theorem \ref{theo:ptopiidachieve2}]
Fix a quantum state $\psi_{AA'}$. Let $n_0 := \frac{n}{b}$, which is assumed to be an integer without loss of generality. Every message $m\in \{1,2,\ldots 2^{nR}\}$ can be decomposed as $m= m_1m_2\ldots m_b$, where each $m_i \in \{1,2,\ldots 2^{n_0R}\}$. 

Alice and Bob share $w$ copies of $\psi_{AA'}$ in the registers $A_1A'_1, \ldots A_wA'_w$.  Additionally, Alice and Bob share $b$ copies of the following randomness in registers $S_1,S_2, \ldots S_{2^{n_0R}}$, 
$$\sum_{s_1, \ldots s_{2^{n_0R}}}\barq(s_1,\ldots s_{2^{n_0R}}) \ketbra{s_1, \ldots s_{2^{n_0R}}}_{S_1, \ldots S_{2^{n_0R}}},$$
where $|S_i| = {w \choose n_0}$, $s_i\in \cS_{n_0,w}$ and $\barq$ is a pairwise independent probability distribution satisfying $\barq(s_i) = \frac{1}{{w \choose n_0}}$ for all $s_i$. 

\noindent {\bf Protocol:} The protocol proceeds in the following rounds. 
\begin{itemize}
\item Set $i=1$. 
\item While  $i\leq b$:
\item Alice and Bob run a $(n_0R, \frac{\eps}{b^3})$ entanglement assisted protocol with their current shared entanglement and current shared randomness, as guaranteed by Theorem \ref{theo:ptopiidachieve}. 
\item  Upon decoding the message $m$, Bob discards the registers $A'_{s_m(1)}, \ldots A'_{s_m(n_0)}$ (the subsets $s_1, s_2, \ldots$ which are used for coding are known to both Alice and Bob, being generated using shared randomness).
\item Alice and Bob consider the remaining quantum registers as their shared entanglement. They invoke a fresh copy of shared randomness for the next round.
\item Set $i \leftarrow i+1$. Go to Step 2.  
\end{itemize}

\noindent {\bf Error Analysis:} Let $\Theta_i$ be the quantum state on the shared entanglement at the beginning of round $i$. We have $\Theta_1 = \psi_{AA'}^{\otimes w}$. For $i>1$, $\Theta_i$ is obtained from $\Theta_{i-1}$ by running above protocol in round $i$ and tracing out the used shared entanglement. Let the resulting quantum map be $\cE_i$. From the gentle measurement lemma (Fact \ref{gentlelemma}) and the fact that the protocol in Theorem \ref{theo:ptopiidachieve} makes an error of at most $\frac{\eps}{b^3}$, we have $$\Pur(\cE_i(\psi^{\otimes w- n_0(i-1)}_{AA'}), \psi^{\otimes w- n_0\cdot i}_{AA'}) \leq \sqrt{\frac{\eps}{b^3}}.$$ Thus, Fact \ref{slowchange} implies that 
$$\Pur(\Theta_i, \psi^{\otimes w- n_0\cdot i}_{AA'}) \leq (i-1)\sqrt{\frac{\eps}{b^3}}.$$
From Lemma \ref{err} we have 
$$\Pr(M'_i \neq m_i \mid  M_i = m_i) \leq (\sqrt{\frac{\eps}{b^3}} + (i-1)\sqrt{\frac{\eps}{b^3}})^2 \leq i^2\frac{\eps}{b^3}.$$
Thus, $$\Pr(M'\neq m \mid M=m) \leq \sum_{i=1}^bi^2\frac{\eps}{b^3} \leq \frac{\eps}{6} \leq \eps.$$

\noindent {\bf Constraints on $R$:} From Theorem \ref{theo:ptopiidachieve}, the protocol can be run as long as for every round $i$, we have 
\begin{eqnarray*}
R &\leq& \frac{1}{n_0}\dmineps{\cN_{A \to B }(\psi_{AA'})^{\otimes n_0}}{\cN_{A\to B}(\psi_{A})^{\otimes n_0} \otimes \psi^{\otimes n_0}_{A'}}{\frac{\eps}{b^3}}\\ &&- \frac{n_0\cdot 2^{\dmax{\psi_{AA'}}{\psi_A\otimes \psi_{A'}}}\log e}{w-n_0\cdot(i-1) - 2n_0} + \frac{1}{n_0}\log \frac{\eps}{b^3}.
\end{eqnarray*}
The choice of $w$ ensures that $w- n_0\cdot (i-1) - 2n_0 \geq \frac{n_0\cdot 2^{\dmax{\psi_{AA'}}{\psi_A\otimes \psi_{A'}}}\log e}{\delta}$, for all $i$. Thus, we can use Fact \ref{dmaxequi} to show that the following constraint on $R$ suffices. 
\begin{eqnarray*}
 R &\leq& \relent{\cN_{A \to B }(\psi_{AA'})}{\cN_{A\to B}(\psi_{A})\otimes \psi_{A'}}- \delta - \frac{1}{n_0}O(\log \frac{n_0 b^3}{\eps}) \\ &&- \sqrt{\frac{1}{n_0}\varrelent{\cN_{A \to B }(\psi_{AA'})}{\cN_{A\to B}(\psi_{A})\otimes \psi_{A'}}}|\Phi^{-1}(\frac{\eps}{b^3})|.
\end{eqnarray*}
Using Fact \ref{gaussianupper}, this is achievable if 
\begin{eqnarray*}
R &\leq& \relent{\cN_{A \to B }(\psi_{AA'})}{\cN_{A\to B}(\psi_{A})\otimes \psi_{A'}}- \delta - \frac{1}{n_0}O(\log \frac{n_0 b^3}{\eps}) \\ &&- \sqrt{\frac{4\log\frac{b^3}{\eps}}{n_0}\varrelent{\cN_{A \to B }(\psi_{AA'})}{\cN_{A\to B}(\psi_{A})\otimes \psi_{A'}}}. 
\end{eqnarray*}
This completes the proof of the theorem.
\end{proof}

An immediate corollary of Theorem \ref{theo:ptopiidachieve2} is the following.
\begin{corollary}
\label{cor:ptopiidrates}
For every pure quantum state $\ket{\psi}_{AA'}$, the following is an achievable pair of communication rate and entanglement rate for the quantum channel $\cN_{A\to B}$:
$$R \leq \mutinf{B}{A'}_{\cN_{A\to B}(\psi_{AA'})}, \quad E \geq S(\psi_A).$$
\end{corollary}
\begin{proof}
Fix $\eps', \delta' \in (0,1)$ and an integer $\ell$ such that the following holds: let $\Pi$ be the projector onto the eigenvectors of $\psi_A^{\otimes \ell}$ with eigenvalues in the range $[2^{-(1+\delta) S(\psi_A)}, 2^{-(1-\delta) S(\psi_A)}]$. Then $\Tr(\Pi\psi_A^{\otimes \ell}) \geq 1-\eps'$. Existence of such an $\ell$ is guaranteed by the Chernoff-Hoeffding bound \cite{Hoeffding63}. Let $\psi'_{A^\ell A'^\ell} : = \frac{\Pi\ketbra{\psi}_{AA'}^{\otimes \ell}\Pi}{\Tr(\Pi\psi_A^{\otimes \ell})}$. 

Let $n$ be large enough such that $n = \ell$. We apply Theorem \ref{theo:ptopiidachieve2} to the quantum state $\psi'_{A^\ell A'^\ell}$ and the channel $\cN_{A\to B}^{\otimes \ell}$, with $b= n^{1/2}$ and $\delta = n^{-\frac{1}{4}}$ (these constants may further be optimized). This gives an $(n\ell \cdot R, \eps)$ entanglement assisted code with 
$$\ell\cdot R \leq \mutinf{B^\ell}{A'^\ell}_{\cN_{A\to B}^{\otimes \ell}\psi'_{A^\ell A'^\ell}} - O(n^{-\frac{1}{4}}).$$ Using Alicki-Fannes inequality (Fact \ref{fact:fannes}),  it suffices to have 
$$R \leq \mutinf{B}{A'}_{\cN_{A\to B}\psi'_{AA'}} - 6\eps'\log|A| - O(n^{-\frac{1}{4}}).$$ 
The rate of shared entanglement is at most $(1 + O(n^{-1/4}))\cdot  (1+\delta') H(\psi_A)$.  
This completes the proof of the corollary.
\end{proof}

\section{Quantum side information about the channel at the encoder}
\label{sec:gelfandpinsker}
\subsection*{Description of task}

\begin{figure}[h]
\centering
\includegraphics[scale=0.4]{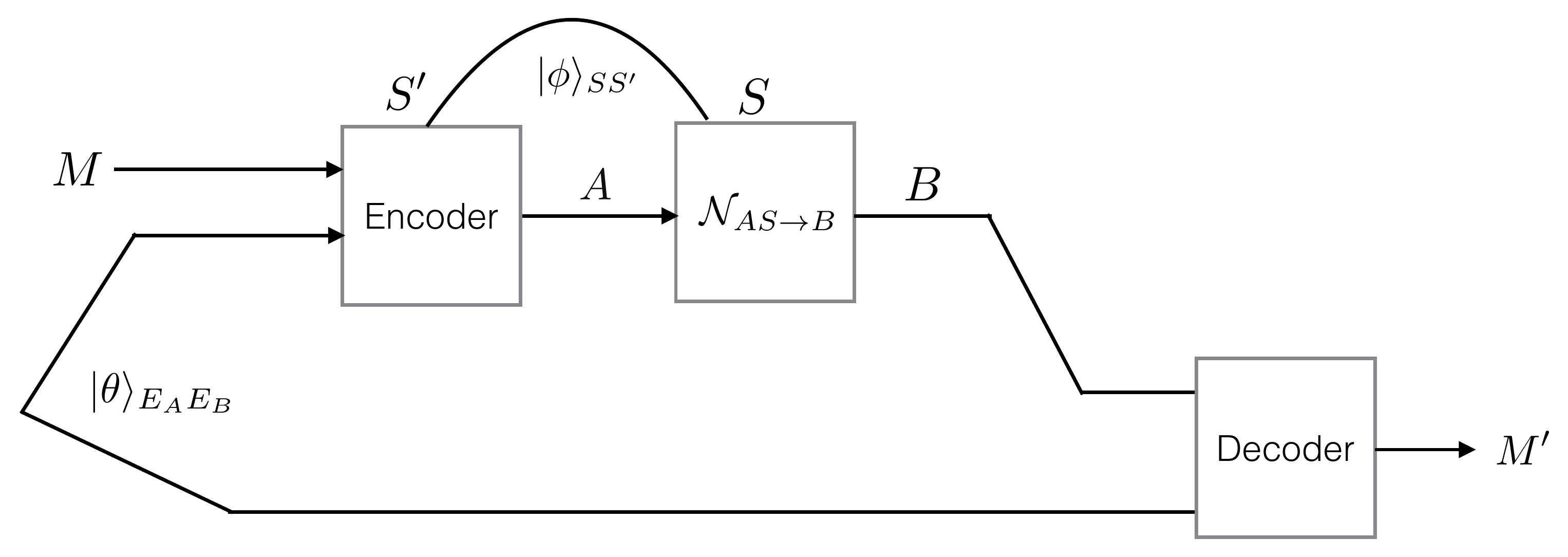}
\caption{ A sketch of general entanglement assisted protocol for point to point channel with quantum side information at the encoder}
\label{fig:sideinfo}
\end{figure}

$\alice$  wants to communicate a classical message $M$ chosen from $[1:2^R]$ to $\bob$  over a quantum channel $\cN_{AS \to B }$ such that $\bob$  is able to decode the correct message with probability at least $1-\eps^2$. $\alice$ shares entanglement with the channel as well. This model in the classical setting is called as the {\em Gel'fand-Pinsker} channel, depicted in Figure~\ref{fig:sideinfo}.
\begin{definition}
\label{codegelpin}
Let $\ket{\theta}_{E_AE_B}$ be the shared entanglement between $\alice$  and $\bob$  and let $\ket{\phi}_{SS'}$ be the state shared between $\alice$  and Channel. An $(R, \eps )$-entanglement assisted code for the quantum channel $\cN_{ AS \to B}$ consists of 
\begin{itemize}
\item An encoding operation $\cE: ME_A S \rightarrow A$ for $\alice$ . 
\item A decoding operation $\cD : B E_B\rightarrow M'$ for $\bob$, with $M'\equiv M$ being the output register such that for all $m$,
\beq
\Pr(M'\neq m|M=m) \leq \eps^2. \nonumber
\enq
\end{itemize}

\end{definition}
\subsection*{An achievability protocol}

\begin{theorem}
Let $\cN_{AS \to B }$ be a quantum channel, $\phi_{SS'}$ be a pure quantum state and let $\eps,\delta \in (0,1)$. Then, for any $R$ satisfying 
\beq
\label{eq:sidechannelrate}
R \leq  \max _{\psi_{AA'S}: \psi_S=\phi_S} \left(\dmineps{\cN_{AS \to B } (\psi_{AA'S})}{\cN(\psi_{AS}) \otimes \psi_{A'}}{\eps} - \imaxeps{\eps}(A':S)_{\psi} \right) - 4\log \frac{1}{\delta},
\enq
there exists an $(R,6\eps+4\delta)$-entanglement assisted code for the quantum channel $\cN_{AS \to B }.$
\end{theorem}

\begin{figure}[h]
\centering
\includegraphics[scale=0.4]{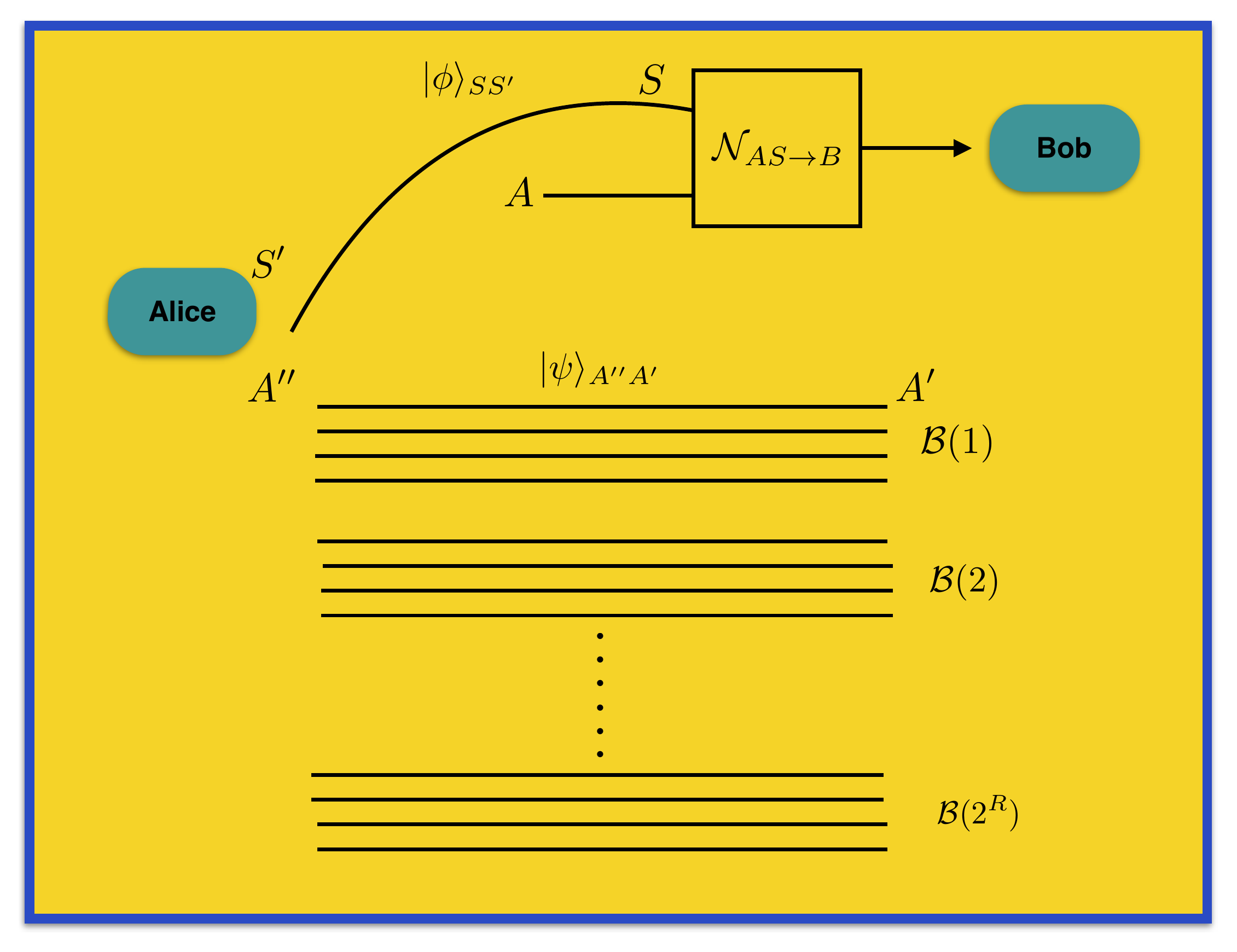}
\caption{A schematic for the achievability protocol. Upon receiving the message $m$, $\alice$ uses the block $\cB(m)$ to transmit the message to $\bob$.}
\label{fig:sideinfoachieve}
\end{figure}

\begin{proof}
Fix $\psi_{AA'S}$ such that $\psi_S=\phi_S$, which achieves the maximum in Equation \ref{eq:sidechannelrate} and fix an $R$ as given in Equation \ref{eq:sidechannelrate}. Set $r = \imaxeps{\eps}(A':S)_{\psi} + 2\log \frac{1}{\delta}.$ 

Let $A''$ be a register such that $\ket{\psi}_{A''A'}$ is a purification of $\psi_{A'}$. Introduce the registers $A''_1,A''_2,\ldots A''_{2^{R + r}}$, such that $A''_i\equiv A''$ and $A'_1,A'_2,\ldots A'_{2^{R+r}}$ such that $A'_i\equiv A'$.  $\alice$  and $\bob$  share the state $$\ketbra{\psi}_{A''_1A'_1}\otimes \ketbra{\psi}_{A''_2A'_2},\ldots \ketbra{\psi}_{A''_{2^{R+r}}A'_{2^{R+r}}},$$
 where $\alice$  holds the registers $A''_1,A''_2, \cdots, A''_{2^{R+r}}$ and $\bob$  holds the registers $A'_1,A'_2, \cdots, A'_{2^{R+r}}$. 

We divide these copies of shared entanglement into blocks, each of size $2^{r}$, where the block $\cB(j)$ involves the registers $A''_{(j-1)\cdot 2^{r}+1}A'_{(j-1)\cdot 2^{r}+1},\ldots A''_{j\cdot 2^{r}}A'_{j\cdot 2^{r}}$ . For brevity, we set $st(j) = (j-1)\cdot 2^{r}+1, en(j)=j\cdot 2^{r}$. Define the following state corresponding to block $j$. 
\begin{equation*}
\tau^j_{SA'_{st(j)}\ldots A'_{en(j)}} \defeq  \frac{1}{2^{r}}\sum_{k\in \cB(j)} \psi_{SA'_k}\otimes\psi_{A'_{st(j)}} \ldots\otimes\psi_{A'_{k-1}}\otimes\psi_{A'_{k+1}}\ldots\otimes\psi_{A'_{en(j)}} .
\end{equation*}
Introduce a register $C$ such that $\ket{\Psi}_{CAA'S}$ is a purification of $\psi_{AA'S}$. Consider the following purification of $\tau^j_{SA'_{st(j)}\ldots A'_{en(j)}}$. 
\begin{align*}
&\ket{\tau^j}_{KACSA''_{st(j)}\ldots A''_{en(j)}A'_{st(j)}\ldots A'_{en(j)}} \\
& = \frac{1}{2^{r/2}}\sum_{k\in \cB(j)}\ket{k}_K \ket{\Psi}_{CASA'_k}\otimes  \ket{\psi}_{A''_{st(j)}A'_{st(j)}} \ldots\otimes\ket{\psi}_{A''_{k-1}A'_{k-1}}\otimes \ket{0}_{A''_k}\otimes\ket{\psi}_{A''_{k+1}A'_{k+1}}\ldots\otimes\ket{\psi}_{A''_{en(j)}A'_{en(j)}} .
\end{align*}
From the corollary of convex split lemma (Corollary \ref{convexcomb}), and the choice of $r$, it holds that
$$\Pur(\tau^j_{SA'_{st(j)}\ldots A'_{en(j)}}, \psi_S\otimes \psi_{A'_{st(j)}}\ldots \otimes \psi_{A'_{en(j)}})\leq 2\eps+\delta.$$ Thus, there exists an isometry $U^j: S' A''_{st(j)}\ldots A''_{en(j)} \rightarrow KACA''_{st(j)}\ldots A''_{en(j)}$ such that (by Uhlmann's Theorem, Fact \ref{uhlmann}) 
\begin{align}
&\Pur(\ketbra{\tau^j}_{KACSA''_{st(j)}\ldots A''_{en(j)}A'_{st(j)}\ldots A'_{en(j)}}, U \ketbra{\psi}_{SS'}\otimes \ketbra{\psi}_{A''_{st(j)}A'_{st(j)}}\otimes\ldots \ketbra{\psi}_{A''_{en(j)}A'_{en(j)}}  U^{\dagger}) \nonumber\\
\label{eq:higfid}
& \leq 2\eps+\delta.
\end{align}
 Our protocol works as follows, also depicted in Figure~\ref{fig:sideinfoachieve}.
 
 \vspace{2mm}
 
\noindent {\bf{Encoding:}} $\alice$  on receiving the message $m \in [1:2^R]$ considers the block $\cB(m)$. She applies the isometry $U^m$ on registers $S'A''_{st(m)}\ldots A''_{en(m)}$ and sends the register $A$ through the channel. Let the state in $\bob$ 's possession after this transmission over the channel be $\hat{\Theta}_{BA'_1\cdots A'_{2^{R}}}$. Define the state
\begin{align*}
\Theta_{BA'_1\cdots A'_{2^{R+r}}}:= \frac{1}{2^{r}}\sum_{k\in \cB(m)} \cN_{ AS \to B}\left( \psi_{ASA'_k}\right) \otimes  \psi_{A'_1} \ldots\otimes\psi_{A'_{k-1}}\otimes\psi_{A'_{k+1}}  \ldots \otimes \psi_{A'_{2^{R+r}}}.
\end{align*}
From Equation \ref{eq:higfid}, and monotonicity of fidelity under quantum operations (Fact \ref{fact:monotonequantumoperation}), it holds that 
\begin{equation}
\label{thetathetahatclose}
\Pur(\Theta_{BA'_1\cdots A'_{2^{R+r}}}, \hat{\Theta}_{BA'_1\cdots,A'_{2^{R+r}}}) \leq 2\eps+\delta.
\end{equation}
{\bf{Decoding:}} 
Let, $ 0 \preceq \Pi_{BA'} \preceq \mathrm{I}$ be such that 
\beq
\label{optimalmeasurement1}
\dmineps{\cN_{AS \to B } (\ketbra{\psi}_{ASA'})}{\cN_{AS\to B}(\psi_{AS}) \otimes \psi_{A'}}{\eps}:= - \log \tr \left[\Pi_{BA'} \cN_{AS\to B}(\psi_{AS}) \otimes \psi_{A'} \right ]. \nonumber
\enq
Using this, we define for each $j \in [1:2^{R_1}]$, 
\begin{equation*} 
\Lambda(j):= \mathrm{I}_{A'_1} \otimes \mathrm{I}_{A'_2} \otimes \cdots \Pi_{BA'_j} \otimes \cdots \otimes \mathrm{I}_{A'_{2^{R+r}}}. 
\end{equation*}
For $j\in [1:2^{R+r}]$, define the operator 
 \begin{equation*}
\Omega(j) := \left(\sum_{ j^\prime \in [1:2^{R+r}]} \Lambda({j'})\right)^{-\frac{1}{2}}\Lambda({j})\left(\sum_{ j^\prime \in [1:2^{R+r}]} \Lambda({j'})\right)^{-\frac{1}{2}}.
\end{equation*}
It is easy to observe that $\sum_j \Omega(j) = \mathrm{I}$, and hence it forms a valid POVM. $\bob$  applies the POVM $\{\Omega(j)\}_{j}$. Upon obtaining the outcome $j$, he outputs the block number corresponding to $j$.

\vspace{2mm}

\noindent {\bf{Probability of error:}} Let $M$ be the message which was transmitted by $\alice$  using the strategy above and let $M'$ be the decoded message by $\bob$  using the above mentioned decoding POVMs. Notice that by the symmetry of the encoding and decoding strategy, it is enough to bound $\Pr \left\{M' \neq 1 \mid  M=1 \right\}$, which we do as follows: 
\begin{align*}
\Pr \left\{M' \neq 1 | M=1\right\} &= \tr \left[\left(\sum_{k' \notin \cB(1)} \Omega(k')\right)\hat{\Theta}_{ A^\prime_1,B_1\cdots,B_{2^{R+r}}}\right]\\
& \overset{a} \leq \left(\sqrt{ \tr \left[\left(\sum_{k' \notin \cB(1)} \Omega(k')\right)\hat{\Theta}_{ A^\prime_1,B_1\cdots,B_{2^{R+r}}}\right]} + \Pur (\Theta, \hat{\Theta})\right)^2\\
& \overset{b} \leq \left(\sqrt{2\eps^2 + 4\delta^2}  + 2\eps+\delta\right)^2 \leq (6\eps+4\delta)^2,
\end{align*}
where $a$ follows from Lemma \ref{err} and $b$ follows from Equation \eqref{thetathetahatclose} and because of the following set of inequalities: 
\begin{align*}
&\tr \left[\left(\sum_{k' \notin \cB(1)} \Omega(k')\right)\Theta_{ B,A^\prime_1\cdots,A^\prime_{2^{R+r}}}\right]\\
 &  \overset{a}  =\frac{1}{2^{r}}\sum_{k\in \cB(1)} \tr \left[\left(\sum_{k' \notin \cB(1)} \Omega(k')\right)  \cN_{ AS \to B}\left( \psi_{ASA'_k}\right) \otimes  \psi_{A'_1} \ldots\otimes\psi_{A'_{k-1}}\otimes\psi_{A'_{k+1}}  \ldots \otimes \psi_{A'_{2^{R+r}}} \right]\\
&\leq   \frac{1}{2^{r}}\sum_{k\in \cB(1)} \tr \left[\left(\sum_{k' \neq k} \Omega(k')\right)  \cN_{ AS \to B}\left( \psi_{ASA'_k}\right) \otimes  \psi_{A'_1} \ldots\otimes\psi_{A'_{k-1}}\otimes\psi_{A'_{k+1}}  \ldots \otimes \psi_{A'_{2^{R+r}}} \right]\\
& \overset{b} =  \tr \left[\left(\sum_{k' \neq 1} \Omega(k')\right)  \cN_{ AS \to B}\left( \psi_{ASA'_1}\right) \otimes  \psi_{A'_2} \ldots \otimes \psi_{A'_{2^{R+r}}} \right]\\
& \overset{c} \leq   2\tr \left[\left(\mathrm{I} -\Lambda(1)\right) \cN_{ AS \to B}\left( \psi_{ASA'_1}\right) \otimes  \psi_{A'_2} \ldots \otimes \psi_{A'_{2^{R+r}}}\right] \\ & \hspace{5mm} + 4\sum_{k^\prime \neq 1} \Tr \left[\Lambda(m) \cN_{ AS \to B}\left( \psi_{ASA'_1}\right) \otimes  \psi_{A'_2} \ldots \otimes \psi_{A'_{2^{R+r}}} \right]\\
&\overset{d} \leq 2 \eps^2 + 4\times 2^{R+r - \dmineps{\cN_{AS \to B } (\ketbra{\psi}_{ASA'})}{\cN_{AS\to B}(\psi_{AS}) \otimes \psi_{A'}}{\eps}} \\
& \overset{e}\leq 2\eps^2 + 4\delta^2 .
\end{align*}
Above $a$ follows from the definition of $\Theta_{ A^\prime_1,B_1\cdots,B_{2^{R+r}}}$; $b$ follows from the symmetry of the code construction; $c$ follows from Hayashi-Nagaoka operator inequality (Fact \ref{haynag}); $d$ follows from the definition of $\Lambda (m)$ and from the definition of $\Pi_{ BA^\prime}$ and $e$ follows from the choice of $R+r$. This completes the proof.
\end{proof}
 
\section{Quantum broadcast channel}
\label{sec:broadcast}

\begin{figure}[h]
\centering
\includegraphics[scale=0.5]{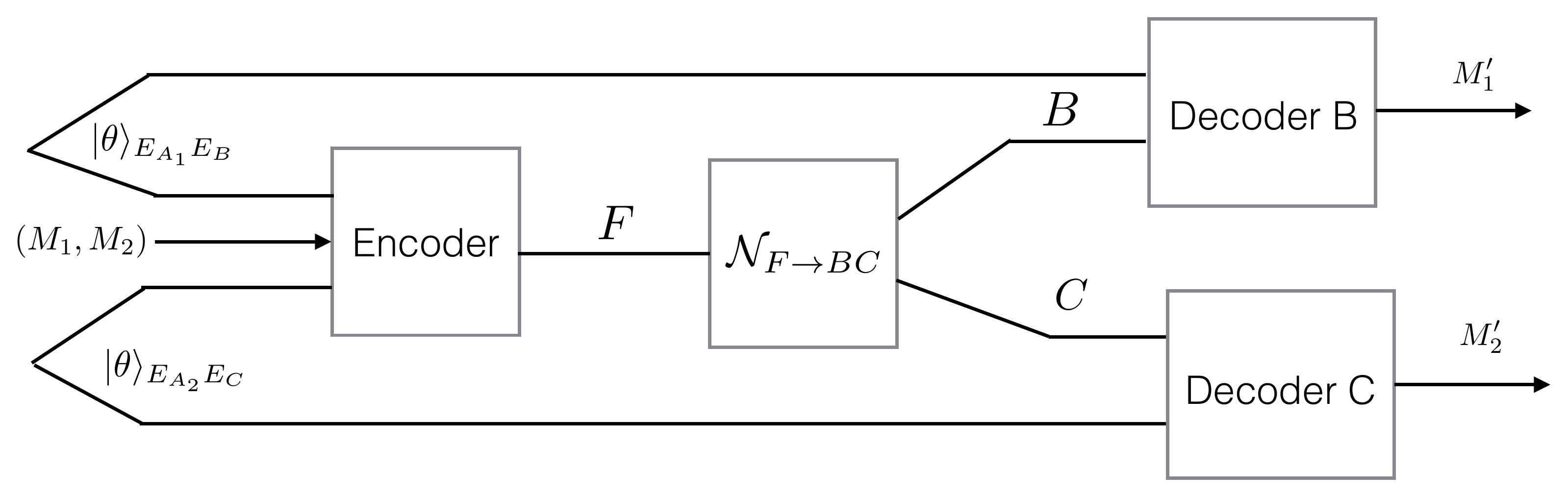}
\caption{Quantum broadcast channel}
\label{fig:quantummarton}
\end{figure}

\subsection*{Description of task:} 

$\alice$ wishes to communicate message pair $(m_1,m_2)$ simultaneously to $\bob$ and $\charlie$ over a quantum broadcast channel, where $m_1$ is intended for $\bob$ and $m_2$ is intended for $\charlie$, such that both $\bob$ and $\charlie$ output the correct message with probability at least $1-\eps^2$. Please refer to Figure~\ref{fig:quantummarton}.
\begin{definition}
\label{broadcastchannelcode}
Let $\ket{\theta}_{E_{A_1}E_B}$ and $\ket{\theta}_{E_{A_2}E_C}$ be the shared entanglement between $\alice$  and $\bob$  and $\alice$  and $\charlie$ respectively. An $(R_1, R_2, \eps )$ entanglement assisted code for the quantum broadcast channel $\cN_{ F \to BC}$ consists of 
\begin{itemize}
\item An encoding operation $\cE: M_1M_2E_{A_1}E_{A_2} \rightarrow F$  for $\alice$ .  
\item A pair of decoding operations $(\cD_B,\cD_C)$,  $\cD_B : B E_B\rightarrow M'_1$ and $\cD_C : C E_C\rightarrow M'_2$, with $(M_1',M_2')\equiv (M_1,M_2)$ being the output registers, such that for all $(m_1,m_2)$
\beq
\Pr((M'_1,M'_2)\neq (m_1,m_2)| (M_1,M_2) = (m_1,m_2))\leq \eps^2 \nonumber
\enq
\end{itemize}

\end{definition}

\subsection*{One-shot Marton inner bound}

Our achievability protocol is inspired by the work\cite{RadhakrishnanSW16}.

\begin{theorem}
\label{oneshotmarton}
Fix $\eps,\delta>0$. Let $\cN:F\rightarrow BC$ be a quantum broadcast channel and let $\psi_{FA_1A_2}$ be a quantum state. Then for any $R_1,R_2$ satisfying 
\begin{eqnarray}
R_1 &\leq& \dmineps{\Tr_C\cN_{F \to BC } (\psi_{FA_1})}{\Tr_C\cN_{F \to BC } (\psi_{F})\otimes \psi_{A_1}}{\eps} - 5\log \frac{1}{\delta} -2, \nonumber\\
R_2 &\leq&  \dmineps{\Tr_B\cN_{F \to BC } (\psi_{FA_2})}{\Tr_B\cN_{F \to BC } (\psi_{F})\otimes \psi_{A_2}}{\eps}  -  5\log \frac{1}{\delta}-2,  \nonumber\\
R_1+R_2 &\leq& \dmineps{\Tr_C\cN_{F \to BC } (\psi_{FA_1})}{\Tr_C\cN_{F \to BC } (\psi_{F})\otimes \psi_{A_1}}{\eps} \nonumber\\ && + \dmineps{\Tr_B\cN_{F \to BC } (\psi_{FA_2})}{\Tr_B\cN_{F \to BC } (\psi_{F})\otimes \psi_{A_2}}{\eps} \nonumber\\ && - \imaxepsbeta{\eps}{\delta}(A_1:A_2)_{\psi}  -  11\log \frac{1}{\delta} - 5,
\end{eqnarray}
the tuple $(R_1,R_2, 4\eps+9\sqrt{\delta})$ is achievable.
\end{theorem}

\begin{figure}[h]
\centering
\includegraphics[scale=0.5]{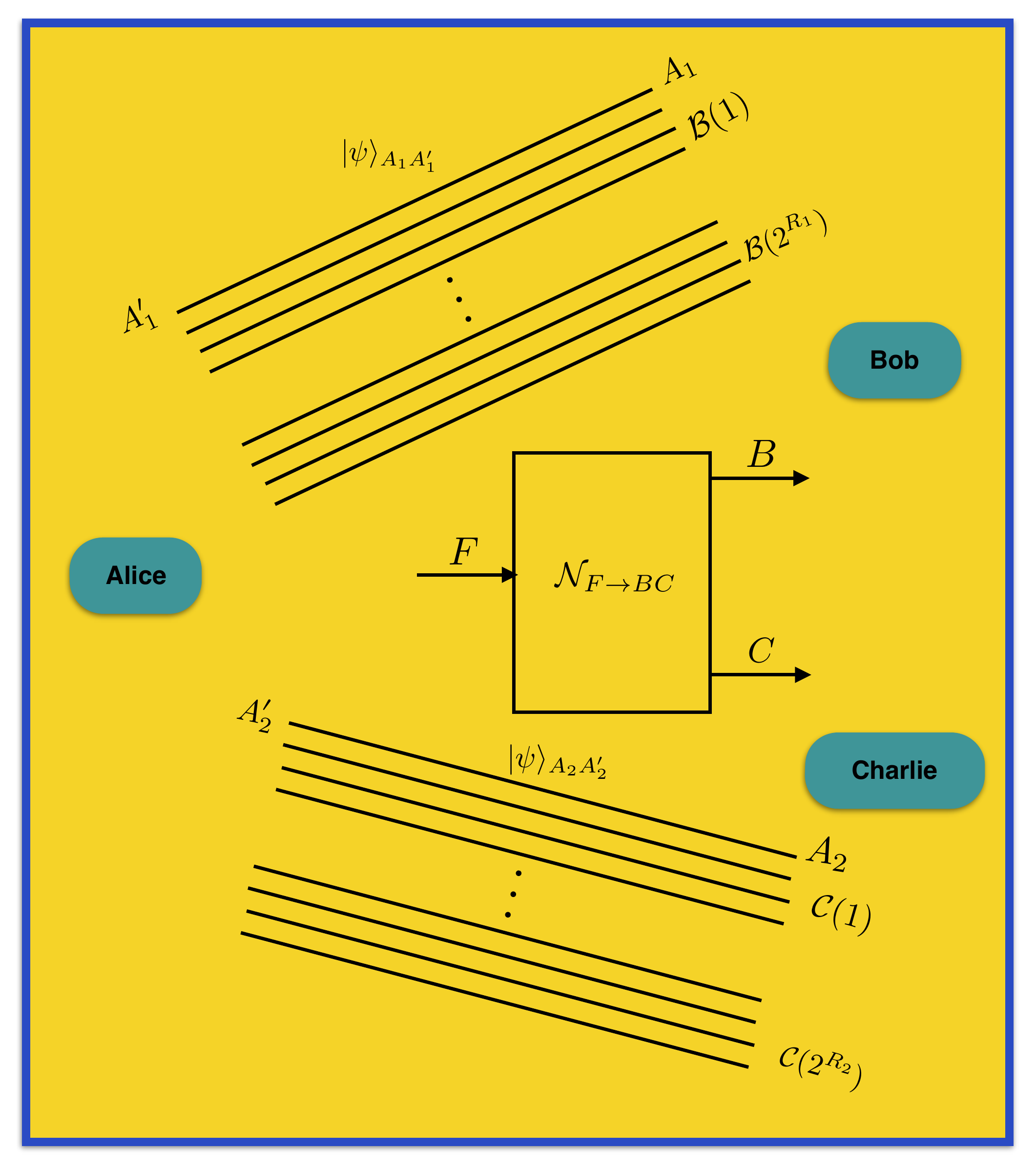}
\caption{Schematic for the achievability proof of quantum broadcast channel.}
\label{fig:quantummartonprotocol}
\end{figure}

\begin{proof}
Given $R_1,R_2$ as in the statement of the theorem, choose $r_1,r_2$  such that $$r_1+r_2 = \lceil\imaxepsbeta{\eps}{\delta}(A_1:A_2)_{\psi} + 3\log\frac{1}{\delta}\rceil, \quad r_1,r_2\geq \log\frac{1}{\delta},$$ $$R_1+r_1 \leq \dmineps{\Tr_C\cN_{F \to BC } (\psi_{FA_1})}{\Tr_C\cN_{F \to BC } (\psi_{F})\otimes \psi_{A_1}}{\eps}  -  4\log \frac{1}{\delta}-1$$ and $$R_2+r_2 \leq \dmineps{\Tr_C\cN_{F \to BC } (\psi_{FA_1})}{\Tr_C\cN_{F \to BC } (\psi_{F})\otimes \psi_{A_1}}{\eps}  -  4\log \frac{1}{\delta}-1.$$
The existence of such $r_1,r_2$ follows from \cite{RadhakrishnanSW16}.

Let $A_1'$ be a register such that $\ket{\psi}_{A_1'A_1}$ is a purification of $\psi_{A_1}$. Introduce the registers $$A'^{(1)},A'^{(2)},\ldots A'^{(2^{R_1+r_1})}$$ such that $A'^{(i)}\equiv A'_1$ and $A^{(1)},A^{(2)},\ldots A^{(2^{R_1+r_1})}$ such that $A^{(i)}\equiv A_1$.  $\alice$  and $\bob$  share the state $$\ketbra{\psi}_{A'^{(1)}A^{(1)}}\otimes \ketbra{\psi}_{A'^{(2)}A^{(2)}},\ldots \ketbra{\psi}_{A'^{(2^{R_1+r_1})}A^{(2^{R_1+r_1})}},$$
 where $\alice$  holds the registers $A'^{(1)},A'^{(2)}, \cdots, A'^{(2^{R_1+r_1})}$ and $\bob$  holds the registers $A^{(1)},A^{(2)}, \cdots,  A^{(2^{R_1+r_1})}$.  We divide these copies of shared entanglement into blocks, each of size $2^{r_1}$, where the block $\cB(j)$ involves the registers $A'^{((j-1)\cdot 2^{r_1}+1)}A^{((j-1)\cdot 2^{r_1}+1)},\ldots A'^{(j\cdot 2^{r_1})}A^{(j\cdot 2^{r_1})}$ . For brevity, we set $st(j) = (j-1)\cdot 2^{r_1}+1, en(j)=j\cdot 2^{r_1}$.

Similarly, let $A_2'$ be a register such that $\ket{\psi}_{A_2'A_2}$ is a purification of $\psi_{A_2}$. Introduce the registers $\hat{A}'^{(1)},\hat{A}'^{(2)},\ldots, \hat{A}'^{(2^{R_2+r_2})}$ such that $\hat{A}'^{(i)}\equiv A'_2$ and $\hat{A}^{(1)},\hat{A}^{(2)},\ldots ,\hat{A}^{(2^{R_2+r_2})}$ such that $\hat{A}^{(i)}\equiv A_2$.  $\alice$  and $\charlie$  share the state $$\ketbra{\psi}_{\hat{A}'^{(1)}\hat{A}^{(1)}}\otimes \ketbra{\psi}_{\hat{A}'^{(2)}\hat{A}^{(2)}},\ldots \ketbra{\psi}_{\hat{A}'^{(2^{R_2+r_2})}\hat{A}^{(2^{R_2+r_2})}},$$
 where $\alice$  holds the registers $\hat{A}'^{(1)},\hat{A}'^{(2)}, \cdots, \hat{A}'^{(2^{R_2+r_2})}$ and $\charlie$  holds the registers $\hat{A}^{(1)},\hat{A}^{(2)}, \cdots, \hat{A}^{2^{(R_2+r_2)}}$. 

We divide these copies of shared entanglement into blocks, each of size $2^{r_2}$, where the block $\cC(j)$ involves the registers $\hat{A}'^{((j-1)\cdot 2^{r_2}+1)}\hat{A}^{(j-1)\cdot 2^{r_2}+1},\ldots \hat{A}'^{(j\cdot 2^{r_2})}\hat{A}^{(j\cdot 2^{r_2})}$ . For brevity, we set $\tilde{st}(j) = (j-1)\cdot 2^{r_2}+1, \tilde{en}(j)=j\cdot 2^{r_2}$.

For the block pair $(\cB(i),\cC(j)),$ where $(i,j) \in [1:2^{R_1}] \times [1:2^{R_2}],$ define the following state:
\begin{align*}
&\tau^{(i,j)}_{A^{st(i)}\ldots A^{en(i)}\hat{A}^{\tilde{st}(j)}\ldots \hat{A}^{\tilde{en}(j)}} \\
&\defeq  \frac{1}{2^{r_1+r_2}}\sum_{(k_1 , k_2 )\in \cB(i) \times \cC(j)} \psi_{A^{(k_1)}\hat{A}^{(k_2)}}\otimes\psi_{A^{(st(i))}} \ldots\otimes\psi_{A^{(k_1-1)}}\otimes\psi_{A^{(k_1+1)}}\ldots\otimes\psi_{A^{(en(i))}}\otimes  \\
& \hspace{65mm}\psi_{\hat{A}^{(st(j))}} \ldots\otimes\psi_{\hat{A}^{(k_2-1)}}\otimes\psi_{\hat{A}^{(k_2+1)}}\ldots\otimes\psi_{\hat{A}^{(\tilde{en}(j))}}.
\end{align*}
Introduce a register $G$ such that $\ket{\Psi}_{GFA_1A_2}$ is a purification of $\psi_{FA_1A_2}$. Now, consider the following purification of $\tau^{(i,j)}_{A^{(st(i))}\ldots A^{(en(i))}\hat{A}^{(\tilde{st}(j))}\ldots \hat{A}^{(\tilde{en}(j))}}$. 
\begin{eqnarray*}
&&\ket{\tau^{(i,j)}}_{KGFA'^{st(i)}\ldots A'^{(en(i))}\hat{A}'^{(\tilde{st}(j))}\ldots \hat{A}'^{(\tilde{en}(j))}A^{(st(i))}\ldots A^{(en(i))}\hat{A}^{(\tilde{st}(j))}\ldots \hat{A}^{(\tilde{en}(j))}} \\ &&\defeq   \frac{1}{2^{(r_1+r_2)/2}}\sum_{(k_1 , k_2 )\in \cB(i) \times \cC(j)}\ket{(k_1,k_2)}_K \ket{\Psi}_{GFA^{k_1}\hat{A}^{k_2}}\otimes \nonumber\\ && \ket{\psi}_{A'^{st(i)}A^{st(i)}} \ldots\otimes\ket{\psi}_{A'^{(k_1-1)}A^{(k_1-1)}}\otimes \ket{0}_{A'^{(k_1)}}\otimes\ket{\psi}_{A'^{(k_1+1)}A^{(k_1+1)}}\ldots\otimes\ket{\psi}_{A'^{(en(i))}A^{(en(i))}}\otimes 
\\ && \ket{\psi}_{\hat{A}'^{(\tilde{st}(i))}\hat{A}^{\tilde{st}(i)}} \ldots\otimes\ket{\psi}_{\hat{A}'^{(k_2-1)}\hat{A}^{(k_2-1)}}\otimes \ket{0}_{\hat{A}'^{(k_2)}}\otimes\ket{\psi}_{\hat{A}'^{(k_2+1)}\hat{A}^{(k_2+1)}}\ldots\otimes\ket{\psi}_{\hat{A}'^{(\tilde{en}(j))}\hat{A}^{(\tilde{en}(j))}}.
\end{eqnarray*}
From Lemma \ref{2dconvexcomb}, and the choice of $r_1,r_2$, it holds that
$$\Pur(\tau^{(i,j)}_{A^{(st(i))}\ldots A^{(en(i))}\hat{A}^{(\tilde{st}(j))}\ldots \hat{A}^{(\tilde{en}(j))}}, \psi_{A^{(st(i))}}\ldots \otimes \psi_{A^{(en(i))}}\otimes \psi_{\hat{A}^{(\hat{st}(i))}}\ldots \otimes \psi_{\hat{A}^{(\hat{en}(i))}})\leq \eps+3\sqrt{\delta}.$$ Thus, there exists an isometry $$U^{(i,j)}: A'^{(st(i))}\ldots A'^{(en(i))}\hat{A}'^{(\tilde{st}(j))}\ldots \hat{A}'^{(\tilde{en}(j))} \rightarrow KGFA'^{(st(i))}\ldots A'^{(en(i))}\hat{A}'^{(\tilde{st}(j))}\ldots \hat{A}'^{(\tilde{en}(j))}$$ such that (by Uhlmann's Theorem) 
\begin{align}
&\Pur\bigg(U\ketbra{\tau^{(i,j)}}U^{\dagger},  \ketbra{\psi}_{A'^{(st(i))}A^{(st(i))}}\otimes\ldots \ketbra{\psi}_{A'^{(en(i))}A^{(en(i))}}\otimes \ketbra{\psi}_{\hat{A}'^{(\tilde{st}(j))}\hat{A}^{(\tilde{st}(j))}}\otimes\ldots \ketbra{\psi}_{\hat{A}'^{(\tilde{en}(j))}\hat{A}^{(\tilde{en}(j))}}\bigg) \nonumber\\
\label{eq:higfidmarton}
& \leq \eps+3\sqrt{\delta}.
\end{align}
Our protocol is as follows. Please refer to Figure~\ref{fig:quantummartonprotocol}.

\vspace{2mm}

\noindent {\bf{Encoding:}} $\alice$  on receiving the message pair $(m_1,m_2 )\in [1:2^{R_1}]\times [1:2^{R_2}]$ considers the block pair $(\cB(m_1),\cC(m_2))$. She applies the isometry $U^{(m_1,m_2)}$ on registers $A'^{(st(i))}\ldots A'^{(en(i))}\hat{A}'^{(\tilde{st}(j))}\ldots \hat{A}'^{(\tilde{en}(j))}$. Then she sends the register $F$ through the channel.

Let the joint  state between the channel output, $\bob$  and $\charlie$ after this transmission over the channel be $$\hat{\Theta}_{BCA^{(1)}\cdots A^{(2^{R_1+r_1})}\hat{A}^{(1)}\cdots \hat{A}^{(2^{R_1+r_1)}}}.$$ Define the state
\begin{align*}
&{\Theta}_{BCA^{(1)}\cdots A^{(2^{R_1+r_1})}\hat{A}^{(1)}\cdots \hat{A}^{(2^{R_1+r_1)}}} \\
&\defeq  \frac{1}{2^{r_1+r_2}}\sum_{(k_1 , k_2 )\in \cB(m_1) \times \cC(m_2)} \cN_{F\rightarrow BC}(\psi_{FA^{(k_1)}\hat{A}^{(k_2)}})\otimes\psi_{A^{(1)}} \ldots\otimes\psi_{A^{(k_1-1)}}\otimes\psi_{A^{(k_1+1)}}\ldots\otimes\psi_{A^{(2^{R_1+r_1)}}} \\
& \hspace{55mm} \otimes\psi_{\hat{A}^{(1)}} \ldots\otimes\psi_{\hat{A}^{(k_2-1)}}\otimes\psi_{\hat{A}^{(k_2+1)}}\ldots\otimes\psi_{\hat{A}^{(2^{R_2+r_2)}}} .
\end{align*}
From Equation \ref{eq:higfidmarton}, and monotonicity of fidelity under quantum operations, it holds that 
\begin{equation}
\label{thetathetahatclosemarton}
\Pur({\Theta}_{BCA^{(1)}\cdots A^{(2^{R_1+r_1})}\hat{A}^{(1)}\cdots \hat{A}^{(2^{R_1+r_1)}}}, \hat{\Theta}_{BCA^{(1)}\cdots A^{(2^{R_1+r_1})}\hat{A}^{(1)}\cdots \hat{A}^{(2^{R_1+r_1)}}}) \leq \eps+3\sqrt{\delta}.
\end{equation}

\vspace{2mm}

\noindent {\bf{Decoding:}} 
Let, $ 0 \preceq \Pi_{BA_1} \preceq \mathrm{I}$ be such that 
\begin{eqnarray*}
&&\dmineps{\Tr_C\cN_{F \to BC } (\psi_{FA_1})}{\Tr_C\cN_{F \to BC } (\psi_{F})\otimes \psi_{A_1}}{\eps} \nonumber\\ && = - \log \tr \left[\Pi_{BA_1}\Tr_C\cN_{F \to BC } (\psi_{F})\otimes \psi_{A_1} \right ].
\end{eqnarray*}
Let, $ 0 \preceq \Pi_{CA_2} \preceq \mathrm{I}$ be such that 
\begin{eqnarray*}
&&\dmineps{\Tr_B\cN_{F \to BC } (\psi_{FA_2})}{\Tr_B\cN_{F \to BC } (\psi_{F})\otimes \psi_{A_2}}{\eps} \nonumber\\ && = - \log \tr \left[\Pi_{CA_2}\Tr_B\cN_{F \to BC } (\psi_{F})\otimes \psi_{A_2} \right ].
\end{eqnarray*}
We begin to define the decoding POVM for $\bob$ . For each $j \in [1:2^{R_1+r_1}]$, define 
\begin{equation*}
\Lambda(j):= \mathrm{I}_{A'^{(1)}} \otimes \mathrm{I}_{A'^{(2)}} \otimes \cdots \Pi_{BA'^{(j)}} \otimes \cdots \otimes \mathrm{I}_{A'^{(2^{R_1+r_1)}}}.
 \end{equation*}
For $j\in [1:2^{R_1+r_1}]$, define the operator 
 \beq
\Omega(j) := \left(\sum_{ j^\prime \in [1:2^{R_1+r_1}]} \Lambda({j'})\right)^{-\frac{1}{2}}\Lambda({j})\left(\sum_{ j^\prime \in [1:2^{R_1+r_1}]} \Lambda({j'})\right)^{-\frac{1}{2}}. \nonumber
\enq
It is easy to observe that $\sum_j \Omega(j) = \mathrm{I}$, and hence it forms a valid POVM. $\bob$  applies the POVM $\{\Omega(j)\}_{j}$. Upon obtaining the outcome $j$, he outputs the block number corresponding to $j$. $\charlie$'s decoding POVM is constructed similarly using the projector $\Pi_{CA_2}$. 

\vspace{2mm}

\noindent {\bf{Probability of error:}} Let $(M_1,M_2)$ be the message pair which was transmitted by $\alice$  using the strategy above and let $(M'_1,M'_2)$ be the decoded message by $\bob$  and $\charlie$ using the above mentioned decoding POVMs. Notice that by the symmetry of the encoding and decoding strategy, it is enough to show that 
$$\Pr \left\{(M'_1,M'_2) \neq (1,1) \mid (M_1,M_2) = (1,1) \right\} \leq  (4\eps+9\sqrt{\delta})^2,$$ under the event that $(M_1,M_2) = (1,1) $ is the transmitted message pair. We upper bound this probability as  follows:
\begin{eqnarray*}
&&\Pr \left\{(M'_1,M'_2) \neq (1,1) \mid (M_1,M_2) = (1,1) \right\} \\ && \leq \Pr \left\{M'_1 \neq 1 \mid (M_1,M_2) = (1,1) \right\} + \Pr \left\{M'_2 \neq 1 \mid (M_1,M_2) = (1,1) \right\} .
\end{eqnarray*}
We now upper bound $\Pr \left\{M'_1 \neq 1 \mid (M_1,M_2) = (1,1) \right\}$ as follows: 
\begin{align*}
\Pr \left\{M' \neq 1 | (M_1,M_2) = (1,1)\right\} &= \tr \left[\left(\sum_{k' \notin \cB(1)} \Omega(k')\right)\hat{\Theta}_{BA^{(1})\cdots A^{(2^{R_1+r_1)}}}\right]\\
& \overset{a} \leq \left(\tr \left[\left(\sum_{k' \notin \cB(1)} \Omega(k')\right){\Theta}_{BA^{(1})\cdots A^{(2^{R_1+r_1)}}}\right]+ \Pur (\Theta, \hat{\Theta})\right)^2\\
& \overset{b} \leq \left(\sqrt{ 2\eps^2 + 8\delta^4}  + \eps+3\sqrt{\delta}\right)^2 \leq (4\eps + 9\sqrt{\delta})^2,
\end{align*}
where $a$ follows from Lemma \ref{err} and $b$ follows from Equation \eqref{thetathetahatclosemarton} and because of the following set of inequalities: 
\begin{align*}
&\tr \left[\left(\sum_{k' \notin \cB(1)} \Omega(k')\right)\Theta_{BA^{(1})\cdots A^{(2^{R_1+r_1)}}}\right]\\
 &  \overset{a}  =\frac{1}{2^{r_1}}\sum_{k\in \cB(1)} \tr \left[\left(\sum_{k' \notin \cB(1)} \Omega(k')\right) \Tr_C\cN_{F\rightarrow BC}(\psi_{FA^{k}})\otimes\psi_{A^{(1)}} \ldots\otimes\psi_{A^{(k_1-1)}}\otimes\psi_{A^{(k_1+1)}}\ldots\otimes\psi_{A^{(2^{R_1+r_1)}}} \right]\\
&\leq   \frac{1}{2^{r_1}}\sum_{k\in \cB(1)} \tr \left[\left(\sum_{k' \neq k} \Omega(k')\right)  \Tr_C\cN_{F\rightarrow BC}(\psi_{FA^{k}})\otimes\psi_{A^{(1)}} \ldots\otimes\psi_{A^{(k_1-1)}}\otimes\psi_{A^{(k_1+1)}}\ldots\otimes\psi_{A^{(2^{R_1+r_1)}}}  \right]\\
& \overset{b} = \tr \left[\left(\sum_{k' \neq 1} \Omega(k')\right)  \Tr_C\cN_{F\rightarrow BC}(\psi_{FA^{(1)}})\otimes\psi_{A^{(2)}} \ldots \otimes\psi_{A^{(2^{R_1+r_1)}}} \right]\\
& \overset{c} \leq   2\tr \left[\left(\mathrm{I} -\Lambda(1)\right)  \Tr_C\cN_{F\rightarrow BC}(\psi_{FA^{(1)}})\otimes\psi_{A^{(2)}} \ldots \otimes\psi_{A^{(2^{R_1+r_1)}}}\right] \\ & \hspace{5mm}+ 4\sum_{k^\prime \neq 1} \Tr \left[\Lambda(m)  \Tr_C\cN_{F\rightarrow BC}(\psi_{FA^{(1)}})\otimes\psi_{A^{(2)}} \ldots \otimes\psi_{A^{2^{(R_1+r_1)}}} \right]\\
&\overset{d} \leq 2 \eps^2 + 4\times 2^{R_1+r_1 - \dmineps{\Tr_C\cN_{F \to BC } (\psi_{FA_1})}{\Tr_C\cN_{F \to BC } (\psi_{F})\otimes \psi_{A_1}}{\eps}} \\
& \overset{e}\leq 2\eps^2 + 8\delta^4 .
\end{align*}
Above $a$ follows from the definition of $\Theta_{ A^\prime_1,B_1\cdots,B_{2^{R}}}$; $b$ follows from the symmetry of the code construction; $c$ follows from Hayashi-Nagaoka operator inequality (Fact \ref{haynag}); $d$ follows from the definition of $\Lambda (m)$ and from the definition of $\Pi_{ BA^\prime}$; and $e$ follows from the choice of $R_1+r_1$. The calculation for $\Pr \left\{M'_2 \neq 1 \mid (M_1,M_2) = (1,1) \right\}$ follows in the similar fashion. This completes the proof. 
\end{proof}

\section{Asymptotic limit of our bounds}
\label{asymptote}

An important property of smooth information theoretic quantities is that in asymptotic setting, they converge to relative entropy based quantities. The achievability bound for point to point channel uses hypothesis testing relative entropy. Fact \ref{dmaxequi} exhibits its asymptotic behaviour.

To exhibit the asymptotic behaviour for quantum channel with side information and rate limited quantum channel with side information (which use an alternative definition of smooth max-information), we shall require the following fact.
\begin{fact}
\label{imaxequi}
Let $\rho^{\otimes n}_{AB}\in \mathcal{D}(A^nB^n)$ be a quantum state. It holds that
\begin{equation*}
\lim_{\eps\to 0}\lim_{n\rightarrow \infty}\frac{1}{n}\imaxepss{\eps}(A:B)_{\rho^{\otimes n}} = \mutinf{A}{B}_{\rho}
\end{equation*}
\end{fact}
\begin{proof}
The proof follows from Corollary B.22 (\cite{Renner11}) and Theorem 3 in \cite{Renner13}.
\end{proof}
Now, the following lemma can be used to show that the quantity $\imaxeps{\eps}(A:B)_{\rho}$ approaches $\mutinf{A}{B}_{\rho}$ in asymptotic and i.i.d setting.
\begin{lemma}
\label{lem:imaxfact}
 For $\rho_{AB}\in \mathcal{D}(AB)$ and $\sigma_A\in \mathcal{D}(A)$, it holds that
$$\imaxeps{2\eps}(A:B)_{\rho} \leq \dmaxeps{\rho_{AB}}{\sigma_A\otimes \rho_B}{\eps} + \log\frac{3}{\eps^2}.$$
\end{lemma}
\begin{proof}
The proof is given in Appendix \ref{append:imaxfact}.
\end{proof}
Finally, we show that the restricted smooth max-information that we have introduced also converges properly. It is crucial in our achievability bound for quantum broadcast channel. We start with the following fact.

\begin{fact}[Lemma 12 and Proposition 13,\cite{TomHay13}]
\label{dmaxepsdseps}
For quantum state $\rho_A\in\mathcal{D}(A)$,  $\sigma \in \mathcal{P}(A)$ and reals $0<\delta <1-\eps^2$, it holds that 
$$ \dseps{\rho_A}{\sigma_A}{1-\eps^2-\delta} - 2\log\frac{1}{\delta}-2 \leq \dmaxeps{\rho_A}{\sigma_A}{\eps} \leq \dseps{\rho_A}{\sigma_A}{1-\eps^2+\delta} + \log v(\sigma) + 2\log\frac{1}{\eps} + \log\frac{1}{\delta},$$
where $v(\sigma_A)$ is the number of distinct eigenvectors of $\sigma_A$. It also holds that 
$$ \dsepsalt{\rho_A}{\sigma_A}{\eps^2+\delta} - 2\log\frac{1}{\delta} -2 \leq \dmaxeps{\rho_A}{\sigma_A}{\eps}.$$
\end{fact}

\begin{proof}
The first part is essentially that given in \cite{TomHay13} (Proposition 13 and Lemma 12). For the second part, we note that the proof in \cite{TomHay13} (Proposition 12, Equation $23$) directly proceeds for this case as well: setting $R\defeq  \dmaxeps{\rho_A}{\sigma_A}{\eps}$, it is shown that for any $\delta'>0$, it holds that $\Tr(\rho_A\{\rho_A- 2^{R+\delta'}\sigma_A\}_{-}) \geq (\sqrt{1-\eps^2}-2^{-\delta'/2})^2$. Setting $\delta' = \log\frac{1}{\delta}$ proves the inequality. 
\end{proof}

We shall also need the well known Chernoff bounds.
\begin{fact}[Chernoff bounds]
\label{fact:chernoff}
Let $X_1 , \ldots, X_n$ be independent random variables, with each $X_i \in [0,1]$ always. Let $X \defeq X_1 + \cdots + X_n$ and  $\mu \defeq \mathbb{E} X = \mathbb{E} X_1 + \cdots + \mathbb{E} X_n$. Then for any $0\leq \varepsilon \leq 1$,
\begin{align*}
\mathrm{Pr}(X \geq (1+\varepsilon)\mu) &\leq \exp\left(-\frac{\varepsilon^2}{3}\mu\right) \\
\mathrm{Pr}(X \leq (1-\varepsilon)\mu) &\leq \exp\left(-\frac{\varepsilon^2}{2}\mu\right)  .
\end{align*}
\end{fact}
Now we proceed to the main results of this section. Using these results and Fact \ref{dmaxequi}, we can easily conclude that for any $\eps < 1/2$, 
$$\lim_{n\rightarrow \infty}\frac{1}{n}\imaxepsbeta{\sqrt{\eps}}{\eps/2}(A^n:B^n)_{\rho} = \mutinf{A}{B}_{\rho}.$$

\begin{theorem}
Let $\rho_{AB}\in \mathcal{D}(AB)$ be a quantum state. Fix an integer $n > 1$ and $\eps < 1/2$ such that: $$n >10^5\cdot\max\{ \frac{\log(1/\lambda_{min}(\rho_A))}{S(\rho_A)\cdot\eps^3}, \frac{\log(1/\lambda_{min}(\rho_B))}{S(\rho_B)\cdot\eps^3} \}.$$ Then it holds that
$$\imaxepsbeta{\sqrt{\eps}}{\eps/2}(A^{\otimes n}:B^{\otimes n})_{\rho^{\otimes n}} \geq \dmaxeps{\rho_{AB}^{\otimes n}}{\rho_{A}^{\otimes n}\otimes \rho_{B}^{\otimes n}}{\sqrt{\eps}}$$
and
 $$\imaxepsbeta{\sqrt{\eps}}{\eps/2}(A^{\otimes n}:B^{\otimes n})_{\rho^{\otimes n}} \leq \dmaxeps{\rho_{AB}^{\otimes n}}{\rho_{A}^{\otimes n}\otimes \rho_{B}^{\otimes n}}{\sqrt{\eps}/24} + 9\log\frac{1}{\eps} + (2|A|+2|B|)\log n + 90.$$
\end{theorem}

\begin{proof}
The first inequality in the statement is trivial. So we consider the second inequality. In below, we will set $\delta = \frac{\eps}{576}$. Our proof is divided into three main steps, as we elucidate below.

\vspace{2mm}

\noindent {\bf Typical projection onto subsystems $A$, $B$:} For brevity, we set $\rho_{A^nB^n} \defeq \rho_{AB}^{\otimes n}$. Let $\Pi_{A^n}$ be the projector onto the eigenvectors of $\rho_{A^n}$ with eigenvalues in the range $[(1-\delta)2^{-n\cdot S(\rho_A)}, (1+\delta)2^{-n\cdot S(\rho_A)}]$. Similarly, let  $\Pi_{B^n}$ be the projector onto the eigenvectors of $\rho_{B^n}$ with eigenvalues in the range $[(1-\delta)2^{-n\cdot S(\rho_B)}, (1+\delta)2^{-n\cdot S(\rho_B)}]$. Let $\mu_{A^n},\mu_{B^n}$ be uniform distributions in the support of $\Pi_{A^n}$ and $\Pi_{B^n}$ respectively. Following relations are easy to observe.
\begin{eqnarray}
\label{uniforms}
&& (1-\delta)\Pi_{A^n}\rho_{A^n}\Pi_{A^n} \leq \mu_{A^n} \leq (1+\delta)\Pi_{A^n}\rho_{A^n}\Pi_{A^n} \leq (1+\delta)\rho_{A^n} \nonumber \\
&& (1-\delta)\Pi_{B^n}\rho_{B^n}\Pi_{B^n} \leq \mu_{B^n} \leq (1+\delta)\Pi_{B^n}\rho_{B^n}\Pi_{B^n} \leq (1+\delta)\rho_{B^n}
\end{eqnarray}
Using Chernoff bounds (Fact \ref{fact:chernoff}),  we have that $\Tr(\Pi_{A^n}\rho_{A^n})\geq 1-2\cdot \exp(-\frac{\delta^2\cdot n\cdot S(\rho_A)}{\log(1/\lambda_{min}(\rho_A))}) \geq 1-\delta$ for the choice of $n$. Similarly, $\Tr(\Pi_{B^n}\rho_{B^n})\geq 1-\delta$.

Now, define the state $\rho'_{A^nB^n} = \frac{(\Pi_{A^n}\otimes \Pi_{B^n})\rho_{A^nB^n}(\Pi_{A^n}\otimes \Pi_{B^n})}{\Tr(\rho_{A^nB^n}(\Pi_{A^n}\otimes \Pi_{B^n}))}$. We will establish the following claims about $\rho'_{A^nB^n}$
\begin{claim}
\label{rhoprimeclaims}
It holds that 
\begin{enumerate}
\item $\F^2(\rho'_{A^nB^n},  \rho_{A^nB^n})\geq 1-16\delta$.
\item $\Tr(\rho_{A^nB^n}(\Pi_{A^n}\otimes \Pi_{B^n})) \geq 1-10\delta$.
\item $\rho'_{A^nB^n} \in \mathrm{supp}(\Pi_{A^n}\otimes \Pi_{B^n})$.
\item $\rho'_{A^n} \leq \frac{1}{1-10\delta}\rho_{A^n}$ and $\rho'_{B^n} \leq \frac{1}{1-10\delta}\rho_{B^n}$.
\end{enumerate}
\end{claim}
\begin{proof}
We prove each item in a sequence below.
\begin{enumerate}
\item This is a straightforward application of gentle measurement lemma (Fact \ref{gentlelemma}). Define the intermediate state $\tau_{A^nB^n} \defeq  \frac{(\id_{A^n}\otimes \Pi_{B^n})\rho_{A^nB^n}(\id_{A^n}\otimes \Pi_{B^n})}{\Tr(\rho_{A^nB^n}(\id_{A^n}\otimes \Pi_{B^n}))}$. From gentle measurement lemma, $\F^2(\tau_{A^nB^n},\rho_{A^n,B^n})\geq \Tr(\Pi_{B^n}\rho_{B^n})\geq 1-\delta$. Moreover, $\Tr(\Pi_{A^n}\rho_{A^n})\geq 1-\delta$. Thus, from Lemma \ref{err}, $\Tr((\Pi_{A^n}\otimes \id_{B^n})\tau_{A^nB^n})\geq 1-9\delta$. Applying gentle measurement lemma again, this gives that $\F^2(\rho'_{A^nB^n},\tau_{A^nB^n})\geq 1-9\delta$. By triangle inequality for purified distance (Fact \ref{fact:trianglepurified}), we conclude that $\F^2(\rho'_{A^nB^n},\rho_{A^nB^n})\geq 1- 16\delta$.
\item As established above, $\Tr((\Pi_{A^n}\otimes \id_{B^n})\tau_{A^nB^n})\geq 1-9\delta$. The item follows by substituting the definition of $\tau_{A^nB^n}$ and using $\Tr(\rho_{B^n}\Pi_{B^n})\geq 1-\delta$.
\item This follows since $(\Pi_{A^n}\otimes \Pi_{B^n})\rho_{A^nB^n}(\Pi_{A^n}\otimes \Pi_{B^n}) < \Pi_{A^n}\otimes \Pi_{B^n}$.
\item We proceed as follows for $\rho'_{B^n}$.  
\begin{eqnarray*}
\rho'_{B^n} &=& \frac{1}{\Tr(\rho_{A^nB^n}(\Pi_{A^n}\otimes \Pi_{B^n}))}\Tr_{A^n}((\Pi_{A^n}\otimes \Pi_{B^n})\rho_{A^nB^n}(\Pi_{A^n}\otimes \Pi_{B^n}))\\ &<& \frac{1}{\Tr(\rho_{A^nB^n}(\Pi_{A^n}\otimes \Pi_{B^n}))}\bigg(\Tr_{A^n}((\Pi_{A^n}\otimes \Pi_{B^n})\rho_{A^nB^n}(\Pi_{A^n}\otimes \Pi_{B^n})) \\ &+& \Tr_{A^n}(((\id_{A^n}-\Pi_{A^n})\otimes \Pi_{B^n})\rho_{A^nB^n}((\id_{A^n}-\Pi_{A^n})\otimes \Pi_{B^n}))\bigg) \\ &=&  \frac{1}{\Tr(\rho_{A^nB^n}(\Pi_{A^n}\otimes \Pi_{B^n}))} (\Tr_{A^n}((\id_{A^n}\otimes \Pi_{B^n})\rho_{A^nB^n}(\id_{A^n}\otimes \Pi_{B^n}))) \\ &=& \frac{1}{\Tr(\rho_{A^nB^n}(\Pi_{A^n}\otimes \Pi_{B^n}))} \Pi_{B^n}\rho_{B^n}\Pi_{B^n} < \frac{1}{1-10\delta} \rho_{B^n}.
\end{eqnarray*}
 Last inequality is due to item $2$ above and the fact that $\Pi_{B^n}$ is a projector onto certain eigenspace of $\rho_{B^n}$. Same argument holds for $\rho'_{A^n}$.
\end{enumerate}
\end{proof}
\noindent{\bf Switching to information spectrum relative entropy:} Using above claim, we now proceed to second step of our proof. As a corollary from the Claim (Item $1$), along with Fact \ref{dmaxepsdseps}, we conclude 
\begin{equation}
\label{switchofdmax}
\dseps{\rho'_{A^nB^n}}{\rho_{A^n}\otimes \rho_{B^n}}{1-50\delta}-2\log\frac{1}{\delta}\leq \dmaxeps{\rho'_{A^nB^n}}{\rho_{A^n}\otimes \rho_{B^n}}{5\sqrt{\delta}} \leq \dmaxeps{\rho_{A^nB^n}}{\rho_{A^n}\otimes \rho_{B^n}}{\sqrt{\delta}} 
\end{equation}
To further simplify this equation, we have the following claim.
\begin{claim}
\label{projecttheprojector}
For any $R>0$, it holds that $$\{\rho'_{A^nB^n} - 2^R\rho_{A^n}\otimes\rho_{B^n}\}_{+} = \{\rho'_{A^nB^n} - 2^R\Pi_{A^n}\rho_{A^n}\Pi_{A^n}\otimes\Pi_{B^n}\rho_{B^n}\Pi_{B^n}\}_{+}.$$
\end{claim}
\begin{proof}
The projector $\Pi_{A^n}$ commutes with $\rho_{A^n}$ and similarly $\Pi_{B^n}$ commutes with $\rho_{B^n}$. For a given $R>0$, consider the operator $O\defeq \rho'_{A^nB^n} - 2^R\rho_{A^n}\otimes\rho_{B^n}$ in the eigenbasis of $\rho_{A^n}\otimes \rho_{B^n}$. Since $\rho'_{A^nB^n}\in \mathrm{supp}(\Pi_{A^n}\otimes \Pi_{B^n})$, which follows from Claim \ref{rhoprimeclaims}, there is no eigenvector of  $O$ orthogonal to the projector $\Pi_{A^n}\otimes \Pi_{B^n}$. Thus, the positive eigenspace of $O$ is equal to the positive eigenspace of $\rho'_{A^nB^n} - 2^R\Pi_{A^n}\rho_{A^n}\Pi_{A^n}\otimes\Pi_{B^n}\rho_{B^n}\Pi_{B^n}$. This proves the claim.
\end{proof}
This claim implies, from the definition of information spectrum relative entropy, that $$\dseps{\rho'_{A^nB^n}}{\rho_{A^n}\otimes \rho_{B^n}}{1-50\delta} = \dseps{\rho'_{A^nB^n}}{\Pi_{A^n}\rho_{A^n}\Pi_{A^n}\otimes \Pi_{B^n}\rho_{B^n}\Pi_{B^n}}{1-50\delta}.$$
Now, we proceed in the following way, setting $v \defeq v(\Pi_{A^n}\rho_{A^n}\Pi_{A^n}\otimes \Pi_{B^n}\rho_{B^n}\Pi_{B^n})$, which is the number of distinct eigenvalues of $\Pi_{A^n}\rho_{A^n}\Pi_{A^n}\otimes \Pi_{B^n}\rho_{B^n}\Pi_{B^n}$:
\begin{eqnarray*}
 &&\dseps{\rho'_{A^nB^n}}{\Pi_{A^n}\rho_{A^n}\Pi_{A^n}\otimes \Pi_{B^n}\rho_{B^n}\Pi_{B^n}}{1-50\delta} \\
&&\overset{a}\geq \dmaxeps{\rho'_{A^nB^n}}{\Pi_{A^n}\rho_{A^n}\Pi_{A^n}\otimes \Pi_{B^n}\rho_{B^n}\Pi_{B^n}}{\sqrt{200\delta}} - \log v - 3\log\frac{1}{\delta}\\ && \overset{b}\geq \dmaxeps{\rho'_{A^nB^n}}{\mu_{A^n}\otimes \mu_{B^n}}{\sqrt{200\delta}} - 2\log\frac{1}{1-\delta} - \log v - 3\log\frac{1}{\delta}\\ && \overset{c}\geq \dsepsalt{\rho'_{A^nB^n}}{\mu_{A^n}\otimes \mu_{B^n}}{400\delta} - 2\log\frac{1}{1-\delta} - \log v - 5\log\frac{1}{\delta}
\end{eqnarray*}
where $(a)$ follows from Fact \ref{dmaxepsdseps}, $(b)$ follows from Fact \ref{triangledmax} and Equation \ref{uniforms} and $(c)$ follows from application of second part of Fact \ref{dmaxepsdseps}.

Combining this with Equation \ref{switchofdmax}, we conclude that
\begin{equation}
\label{uniformandactual}
\dsepsalt{\rho'_{A^nB^n}}{\mu_{A^n}\otimes \mu_{B^n}}{400\delta}\leq \dmaxeps{\rho_{A^nB^n}}{\rho_{A^n}\otimes \rho_{B^n}}{\sqrt{\delta}} + 8\log\frac{1}{\delta} + \log v.
\end{equation}
{\bf Removing large eigenvalues:} Now we are in a position to proceed through the final step. Let $R'$ be the minimum achieved in $\dsepsalt{\rho'_{A^nB^n}}{\mu_{A^n}\otimes \mu_{B^n}}{400\delta}$. For brevity, set $\Pi' \defeq \{\rho'_{A^nB^n}-2^{R'}\mu_{A^n}\otimes \mu_{B^n}\}_{-}$ and  define the state $\rho''_{A^nB^n}\defeq \frac{\Pi'\rho'_{A^nB^n}\Pi'}{\Tr(\Pi'\rho'_{A^nB^n})}$. It holds that \item $\Tr(\Pi'\rho'_{A^nB^n}) \geq 1 - 400\delta$. We prove the following properties for $\rho''_{A^nB^n}$.
\begin{claim}
\label{rhodoubleprimeclaims} It holds that
\begin{enumerate}
\item $\Pur(\rho''_{A^nB^n},\rho_{A^nB^n})\leq 24\sqrt{\delta}$.
\item $\rho''_{A^n} < (1+1000\delta)\rho_{A^n}, \rho''_{B^n} < (1+1000\delta)\rho_{B^n}$
\item $ \dmax{\rho''_{A^nB^n}}{\rho_{A^n}\otimes \rho_{B^n}} \leq \dmaxeps{\rho_{A^nB^n}}{\rho_{A^n}\otimes \rho_{B^n}}{\sqrt{\delta}} + 9\log\frac{1}{\delta} + \log v$.
\end{enumerate}
\end{claim}

\begin{proof}
We prove the items in the respective sequence.
\begin{enumerate}

\item From gentle measurement lemma \ref{gentlelemma}, we have that $\F^2(\rho''_{A^nB^n},\rho'_{A^nB^n})\geq \Tr(\Pi'\rho'_{A^nB^n}) \geq 1-400\delta$. Using Claim \ref{rhoprimeclaims} and triangle inequality for purified distance (Fact \ref{fact:trianglepurified}), we obtain that $\Pur(\rho''_{A^nB^n},\rho_{A^nB^n})\leq 24\sqrt{\delta}$. 

\item  Since $\mu_{A^n}\otimes \mu_{B^n}$ is uniform in the support of $\rho'_{A^nB^n}$, $\rho'_{A^nB^n}$ commutes with $\mu_{A^n}\otimes \mu_{B^n}$. This immediately implies that $\Pi'$ commutes with $\rho'_{A^nB^n}$. Thus, we conclude that $$\rho''_{A^nB^n} = \frac{\Pi'\rho'_{A^nB^n}\Pi'}{\Tr(\Pi'\rho'_{A^nB^n})} \leq \frac{\rho'_{A^nB^n}}{\Tr(\Pi'\rho'_{A^nB^n})}\leq \frac{\rho'_{A^nB^n}}{1-400\delta} \leq \frac{\rho'_{A^nB^n}}{1-400\delta},$$ where the second last inequality follows from the relation $\Tr(\Pi'\rho'_{A^nB^n})\geq 1-400\delta$.

Invoking Claim \ref{rhoprimeclaims}, we obtain $$\rho''_{A^n} \leq \frac{\rho'_{A^n}}{1-400\delta} \leq \frac{\rho_{A^n}}{(1-400\delta)(1-10\delta)} \leq \frac{\rho_{A^n}}{1-410\delta}.$$ Similarly, we obtain $\rho''_{B^n} \leq \frac{\rho_{B^n}}{1-410\delta}$. The item now follows since $\frac{1}{1-410\delta} < 1+1000\delta$ for the choice of $\delta$.

\item By definition of $\Pi'$, we have that $\Pi'\tau_{A^nB^n}\Pi' \leq 2^{R'}\Pi'\mu_{A^n}\otimes \mu_{B^n}\Pi' \leq 2^{R'}\mu_{A^n}\otimes \mu_{B^n}$, where last inequality holds since $\mu_{A^n}\otimes \mu_{B^n}$ is uniform. Thus, $\rho''_{A^nB^n} < \frac{2^{R'}}{1-410\delta}\cdot\mu_{A^n}\otimes\mu_{B^n}.$

From Equation \ref{uniforms}, this further implies that
$$\rho''_{A^nB^n} < \frac{(1+\delta)^2\cdot 2^{R'}}{1-410\delta}\cdot\rho_{A^n}\otimes\rho_{B^n}.$$ This proves the item after using Equation \ref{uniformandactual} to upper bound $R'$.
\end{enumerate}
\end{proof}

This claim allows us to conclude that $\rho''_{A^nB^n}$ forms a feasible solution for the optimization in $\imaxepsbeta{24\sqrt{\delta}}{1000\delta}(A^n:B^n)_{\rho}$. 

Now the value of $v$, which is the number of distinct eigenvalues of $\Pi_{A^n}\rho_{A^n}\Pi_{A^n}\otimes \Pi_{B^n}\rho_{B^n}\Pi_{B^n}$, is upper bounded by the number of distinct eigenvalues of $\rho_{A^n}\otimes \rho_{B^n}$. This is at most $n^{2|A|+2|B|}$. This proves the theorem.
\end{proof}

\subsection*{Conclusion}
To summarize, our work exhibits that the techniques of convex split and position based decoding are sufficient to design protocols (similar in spirit to their classical counterparts) for noisy quantum networks. Moreover, these techniques allow us to obtain optimal bounds for communication over entanglement assisted point-to-point quantum channel.

\begin{figure}[h]
\centering
\includegraphics[scale=0.25]{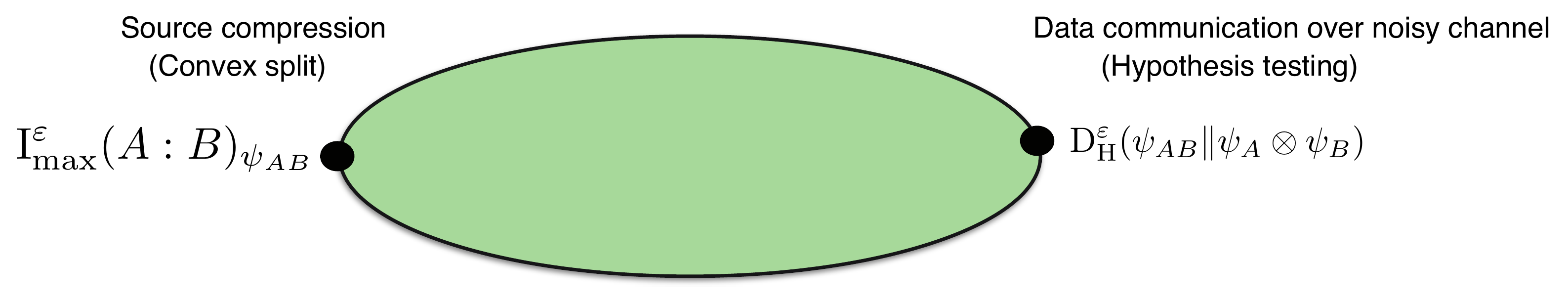}
\caption{{Two fundamental quantities of one-shot quantum information theory}}
\label{fig:tomcoverellipse}
\end{figure}

In the classical asymptotic setting, the well known book on information theory by Thomas and Cover \cite[Figure 2.1]{CoverT91} highlights that there are two fundamental quantities in information theory $\max_{p_X} \mutinf{X}{Y}$ and $\min_{p_{Y|X}} \mutinf{X}{Y}$, each relevant in the contexts of channel and source coding respectively. In the same spirit, our work highlights that there are two fundamental quantities in one-shot (classical and) quantum information theory, smooth hypothesis testing divergence and smooth max R{\'e}nyi divergence (Figure~\ref{fig:tomcoverellipse}, inspired from \cite[Figure 2.1]{CoverT91}, captures this perspective). This is further strengthened by a series of recent works~\cite{AnshuJW17SR,AnshuJW17MC,AnshuJW17SW,Wil17a,Wil17b} which obtain bounds for several different quantum communication tasks in terms of either smooth hypothesis testing divergence or smooth max R{\'e}nyi divergence or both, all using the techniques of convex split and position based decoding. 

\subsection*{Acknowledgement}
This work is supported by the Singapore Ministry of Education and the National Research Foundation, through the Tier 3 Grant ``Random numbers from quantum processes'' MOE2012-T3-1-009 and NRF RF Award NRF-NRFF 2013-13.

\bibliographystyle{ieeetr}
\bibliography{References}

\appendix 

\section{Proof of Lemma \ref{lem:imaxfact}}
\label{append:imaxfact}
\begin{proof}
The proof follows closely from \cite{Renner13}. We partially reproduce it here for completeness. 
We begin with the following claim. 
\begin{claim}
\label{rennerclaim}
For quantum states $\sigma_A,\sigma_B,\psi_{AB}$, there exists a state $\bar{\psi}_{AB} \in \ball{\eps}{\psi_{AB}}$, such that
$$\dmax{\bar{\psi}_{AB}}{\bar{\psi}_A\otimes \sigma_B} \leq \dmax{\psi_{AB}}{\sigma_A\otimes \sigma_B} + \log\frac{3}{\eps^2}.$$
\end{claim}
\begin{proof}
The proof as given in \cite{Renner13} is as follows. Define the operator $\Gamma_A\defeq \psi_A^{\frac{-1}{2}}\sigma_A\psi_A^{-\frac{1}{2}}$ and let $\Pi_A$ be the minimum rank projector such that $\Tr(\Pi_A\psi_A)\geq \sqrt{1-\eps^2}$. Then, it is shown that $\|\Pi_A\Gamma_A\Pi_A\|_{\infty}\leq \frac{1}{1-\sqrt{1-\eps^2}} \leq \frac{2}{\eps^2}$.

Now, chosing $\psi_{ABC}$ as a purification of $\psi_{AB}$, $\Pi_A$ is used to construct a projector $\Pi_{BC}$, which is then used to define the operator $\psi'_{ABC}\defeq \Pi_{BC}\psi_{ABC}\Pi_{BC}$. It is shown that $\psi'_{ABC} \in \ball{\eps}{\psi_{ABC}}$. Now, 
\begin{eqnarray*}
2^{\dmax{\psi'_{AB}}{\psi_A\otimes\sigma_B}} &=& \|\psi^{-1/2}_A\otimes\sigma^{-1/2}_B \psi'_{AB} \psi^{-1/2}_A\otimes\sigma^{-1/2}_B\|_{\infty}\\ &\leq& 2^{\dmax{\psi_{AB}}{\sigma_A\otimes\sigma_B}}\|\Pi_A\Gamma_A\Pi_A\|_{\infty} \leq 2^{\dmax{\psi_{AB}}{\sigma_A\otimes\sigma_B}}\cdot \frac{2}{\eps^2}.
\end{eqnarray*}
Now defining the operator $\psi''_{AB} \defeq \psi'_{AB} + (\psi_A-\psi'_A)\otimes\sigma_B$, which satisfies $\psi''_{AB}\in \ball{\eps}{\psi_{AB}}$, one computes 
\begin{eqnarray*}
2^{\dmax{\psi''_{AB}}{\psi''_A\otimes\sigma_B}} &=& \|\psi''^{-1/2}_A\otimes\sigma^{-1/2}_B \psi''_{AB} \psi''^{-1/2}_A\otimes\sigma^{-1/2}_B\|_{\infty} \\ & = & \|\psi^{-1/2}_A\otimes\sigma^{-1/2}_B \psi''_{AB} \psi^{-1/2}_A\otimes\sigma^{-1/2}_B\|_{\infty}\\ &\leq& \|\psi^{-1/2}_A\otimes\sigma^{-1/2}_B \psi'_{AB} \psi^{-1/2}_A\otimes\sigma^{-1/2}_B\|_{\infty} + 1 \\ &\leq& 2^{\dmax{\psi_{AB}}{\sigma_A\otimes\sigma_B}}\cdot \frac{2}{\eps^2} + 1.
\end{eqnarray*}
Choosing $\frac{\psi''_{AB}}{\Tr(\psi''_{AB})}$ as the desired state, the claim follows.   
\end{proof}
Using this claim, we proceed as follows. Let $\rho'_{AB}\in \ball{\eps}{\rho_{AB}}$ be the quantum state that achieves the minimum in the definition of $\dmaxeps{\rho_{AB}}{\sigma_A\otimes \rho_B}{\eps}$. Using Claim \ref{rennerclaim}, there exists a state $\rho''_{AB}\in \ball{\eps}{\rho'_{AB}}\in\ball{2\eps}{\rho_{AB}}$ such that 
$$\dmax{\rho''_{AB}}{\rho''_A\otimes \rho_B} \leq \dmax{\rho'_{AB}}{\sigma_A\otimes \rho_B} + \log\frac{3}{\eps^2}.$$
This proves the lemma.
\end{proof}

\end{document}